\documentclass[aps,reprint,pra,longbibliography,groupedaddress]{revtex4-2}
\usepackage{}
\usepackage{amssymb}
\usepackage{amsmath}
\usepackage{dcolumn}
\usepackage{bm}
\usepackage{graphicx}
\usepackage{mathrsfs}
\usepackage[colorlinks,linkcolor=blue,anchorcolor=blue,citecolor=blue,urlcolor=blue]{hyperref}
\usepackage{color}
\usepackage{amsmath}

\begin{document}

\preprint{APS/123-QED}

\title{ Collective radiance with NV centers coupled to nonlinear phononic waveguides}

\author{Jia-Qiang Chen}
\affiliation{Ministry of Education Key Laboratory for Nonequilibrium Synthesis and Modulation of Condensed Matter, Shaanxi Province Key Laboratory of Quantum Information and Quantum Optoelectronic Devices, School of Physics, Xi'an Jiaotong University, Xi'an 710049, China }
\author{Yi-Fan Qiao}
\affiliation{Ministry of Education Key Laboratory for Nonequilibrium Synthesis and Modulation of Condensed Matter, Shaanxi Province Key Laboratory of Quantum Information and Quantum Optoelectronic Devices, School of Physics, Xi'an Jiaotong University, Xi'an 710049, China }
\author{Xing-Liang Dong}
\affiliation{Ministry of Education Key Laboratory for Nonequilibrium Synthesis and Modulation of Condensed Matter, Shaanxi Province Key Laboratory of Quantum Information and Quantum Optoelectronic Devices, School of Physics, Xi'an Jiaotong University, Xi'an 710049, China }
\author{Cai-Peng Shen}
\affiliation{Ministry of Education Key Laboratory for Nonequilibrium Synthesis and Modulation of Condensed Matter, Shaanxi Province Key Laboratory of Quantum Information and Quantum Optoelectronic Devices, School of Physics, Xi'an Jiaotong University, Xi'an 710049, China }
\author{Xin-Lei Hei}
\affiliation{Ministry of Education Key Laboratory for Nonequilibrium Synthesis and Modulation of Condensed Matter, Shaanxi Province Key Laboratory of Quantum Information and Quantum Optoelectronic Devices, School of Physics, Xi'an Jiaotong University, Xi'an 710049, China }
\author{Peng-Bo Li}
 \email{lipengbo@mail.xjtu.edu.cn}
\affiliation{Ministry of Education Key Laboratory for Nonequilibrium Synthesis and Modulation of Condensed Matter, Shaanxi Province Key Laboratory of Quantum Information and Quantum Optoelectronic Devices, School of Physics, Xi'an Jiaotong University, Xi'an 710049, China }


\begin{abstract}
Collective radiance is a fundamental phenomenon in quantum optics. However, these  radiation effects remain largely unexplored
in the field of quantum acoustics.
In this work, we investigate the  supercorrelated radiation effects  in a nonlinear phononic waveguide that is coupled with NV centers.
When the spin's frequency is below the scattering continuum but within the bound-state band of the phonon waveguide, a single NV center dissipates slowly, but two NV centers can exhibit a rapid exponential decay.
When multiple NV spins are considered, supercorrelated radiance occurs at a rate N times faster than Dicke superradiance.
The peak of the state distribution in supercorrelated radiance jumps directly from $|m=N/2\rangle$ to $|m=-N/2\rangle$, distinguished from the continuous shift of the peak in superradiance.
This work provides deeper insight into the collective radiation effect  and may find interesting applications in quantum information processing.

%\begin{description}
%42.50.Dv, 03.67.-a, 76.30.Mi
%\end{description}
\end{abstract}

\maketitle

%\tableofcontents

\section{introduction}
The cooperative spontaneous radiance has always been a topic of interest in quantum optics.
In 1954, Dicke first proposed the concept of superradiance \cite{PhysRev.93.99,GROSS1982301,PhysRevA.3.1735,PhysRevLett.118.123602} to describe the enhanced emissions of photons from multiple closely spaced atoms.
Later the concept of subradiance \cite{PhysRevLett.54.1917,PhysRevLett.76.2049,PhysRevLett.116.083601} was developed and described a suppressed cooperative emission, which is slower than incoherent emission.
Superradiance can be used for fast writing and light source due to fast emission rate \cite{PhysRevLett.115.243602,Kovalev_2018,bin2021parity}, while subradiance can enhance the coherence time of quantum states and quantum memory \cite{PhysRevX.7.031024,PhysRevLett.126.103604,PhysRevLett.122.093601,PhysRevLett.117.243601,PhysRevResearch.2.023086}.
Both superradiance and subradiance are derived from interference effects of an atomic ensemble in the environment consisting of independent harmonic oscillators.
Since natural nonlinearities tend to be very weak, the discussion in non-interacting baths is usually valid.
However, it is necessary and interesting to explore cooperative emission in special or artificially designed environments with considerable nonlinearity.

On the other hand, the successful application of laser cooling techniques to cool isolated vibrational modes of nanomechanical devices \cite{Arcizet2006,Schliesser2008,wilson2008cavity,Rocheleau2010,o2010quantum,lai2021domino} has recently led to much interest in quantum acoustics.
With the great progress of nanofabrication techniques, the diamond nanomechanical resonator (NAMR) has achieved quality factor exceeding one million ($Q>10^6$) \cite{Tao2014,doi:10.1063/1.4760274,doi:10.1063/1.2804573}.
And the coherent interfaces between mechanics and other systems such as photons, superconducting qubits, ultra-cold atoms, quantum dots, and lattice defects have been implemented experimentally \cite{RevModPhys.86.1391,Kippenberg1172,LaHaye2009,Jockel2015,doi:10.1021/nl501413t,Yeo2014,Kolkowitz1603}.
In analogy to photons, phonons can be confined in mechanical resonator, as well as propagating freely along phononic waveguides.
As the carrier of information, the moving phonon provides a quantum channel for information exchange between different nodes, which can be used to implement scaling quantum networks \cite{PhysRevX.8.041027,PhysRevLett.120.213603,PhysRevX.5.031011,Lemonde_2019}.
Besides, schemes for producing strong phonon-phonon interactions have been proposed, which can be used to achieve phonon blockade \cite{PhysRevA.82.032101,PhysRevLett.121.153601,PhysRevA.99.013804,PhysRevA.93.063861} and single phonon source \cite{Cai:18}.

Recently, lattice defects in diamond like nitrogen vacancy (NV) centers are particularly attractive  \cite{PhysRevLett.112.047601,PhysRevLett.117.015502,PhysRevX.6.041060,PhysRevLett.116.143602,PhysRevA.96.062333,PhysRevApplied.10.024011,PhysRevApplied.4.044003,ai2021nv,xiong2021strong,hei2021enhancing,zhou2021chiral,huillery2020spin}.
The NV center exhibits long coherence time even at room temperature ($T_2\sim \text{ms}$) \cite{Bar-Gill2013,Balasubramanian2009,Maurer1283,doi:10.1063/1.1626791,PhysRevLett.101.047601,Du2009,Balasubramanian2009}, and its spin and orbital degrees of freedom have been precisely controlled experimentally \cite{Bassett1333,Buckley1212,PhysRevB.81.035205,PhysRevX.2.031001,PhysRevLett.97.083002,PhysRevLett.110.167402,Gao2015}.
Since  NV centers are sensitive to magnetic field and strain field, and can not only be indirectly coupled to NAMRs through magnetic tip \cite{PhysRevLett.125.153602,PhysRevB.79.041302,PhysRevA.98.052346,Arcizet2011}, but also directly coupled by lattice strain \cite{PhysRevLett.110.156402,PhysRevLett.113.020503,Ovartchaiyapong2014,li2020simulation,li2021simulation}.
A spin-mechanical system based on direct coupling may be more attractive because it avoids the emergence of extra decoherence by more components.
Such simple spin-mechanical systems not only provide a promising platform for studying fundamental physics \cite{dong2021unconventional}, but also have a wide variety of applications, ranging from quantum information processing to quantum sensing.

In this work, we propose a scheme to investigate the collective radiance of NV centers in a nonlinear phononic waveguide consisting of coupled NAMRs with Kerr-like nonlinearities.
We first discuss the band structure of the waveguide, which not only contains scattering continuums in regular waveguides, but also has a novel bound-state band.
The bound state consists of a dimer of two phonons which are close to each other and can propagate freely in the waveguide.
When the frequency of NV spin is below the scattering continuum, it can still be located in the bound-state band, which induces two-phonon decay.
We found that a single NV center dissipates slowly, while two NV centers exhibit a rapid exponential decay induced by the novel radiance channel.
By adjusting the position and frequency of the NV center, we achieve subradiance and supercorrelated radiance which is about $N$ times faster than Dicke superradiance.
The peak of the state distribution in supercorrelated radiance jumps directly from $|m=N/2\rangle$ to $|m=-N/2\rangle$,  distinguished from the continuous shift of the peak in superradiance.
Finally, we reproduce the superradiance effect by using the two-phonon decay in this nonlinear waveguide.
Although the dissipative channel and the way of state evolution are completely different from the superradiance situation, the essence of collective radiance in a nonlinear environment is still interference.
This work deepens the insight into collective radiance and may find various applications in  quantum information processing.

\section{The Model}
\begin{figure}
	\includegraphics[width=8cm]{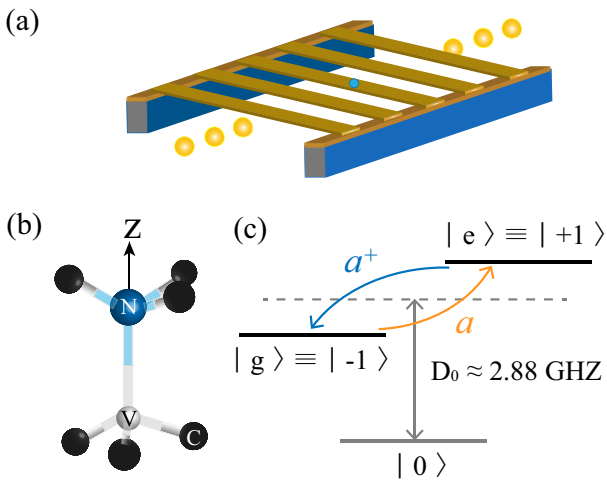}
	\caption{\label{fig1}(Color online)
		(a) Sketch of the phononic waveguide.
		The NAMRs exhibit the nearest neighbor interaction $J$ and the in-site attractive interaction $U$.
		The NV centers are coupled to the NAMRs via lattice strain.
		(b) Molecular structures of the NV center.
		(c) Spin triplet states of the NV electronic ground state.
		Local strain induced by beam bending mixes the states $|\pm1\rangle$.
	}
\end{figure}
As illustrated in Fig.~\ref{fig1}(a), the setup under consideration is an array of coupled NAMRs with Kerr-like nonlinearities, embedded with NV centers.
The NAMR is made of a doubly clamped diamond beam with dimensions $L\gg w, t$, and thus can be described by Euler-Bernoulli thin beam elasticity theory. In this case, the phononic waveguide can be described by the following Hamiltonian
\begin{eqnarray}\label{eq1}
H_r=\sum_n^{N_r}\omega_r a^\dag_n a_n -J(a^\dag_na_{n+1}+a_na^\dag_{n+1})
- \frac{U}{2} a^\dag_n a^\dag_n a_n a_n, \notag\\
\end{eqnarray}
where $a_n(a^\dag_n)$ and $\omega_r$ are the annihilation (creation) operators and frequency for resonator at the site $n$, respectively.
$N_r$ is the number of resonators.
The second term denotes  tunneling couplings between nearest-neighbor resonators with the amplitude $J>0$.
And the last term describes a Kerr-like nonlinear interaction with the amplitude $U>0$.
The basic physics of interacting phonons in a lattice, is contained in the Bose-Hubbard model with a on-site attractive interaction \cite{Winkler2006}.
The intrinsic nonlinearity of the mechanical resonator is generally weak while a strong nonlinearity $U\sim J$ is required in our scheme.
The most common approach is to couple the mechanical mode to an auxiliary system, such as a superconducting qubit \cite{lu2015steady}.
In addition, softening the mechanical mode can also enhance the resonator's nonlinearity \cite{lu2013quantum,Rips_2012}.

The ground state of the NV center is spin $S=1$ triplet states, labeled as $|m_s\rangle$ with $m_s=0,\,\pm1$ (Fig.~\ref{fig1}(c)).
The nonaveraged spin-spin interaction leads to a zero-field splitting $D_0/2\pi\simeq2.88$ GHz between $|0\rangle$ and $|\pm1\rangle$ \cite{PhysRevB.85.205203,DOHERTY20131}.
In the presence of a static magnetic field $\vec{B}=B\vec{e}_z$, the degenerate spin states $|\pm1\rangle$ are split into $|g\rangle\equiv|-1\rangle$ and $|e\rangle\equiv|+1\rangle$ with energy difference $\omega_e=\gamma_SB$, where $\gamma_S$ is the spin gyromagnetic ratio.
When the resonator vibrates, the local strain field changes (resulting in an effective electric field), which results in splitting and mixing of the spin levels $|g\rangle$ and $|e\rangle$ \cite{PhysRevLett.113.020503,PhysRevLett.110.156402,Lee_2017}.
Therefore in the subspace $\{|g\rangle,|e\rangle\}$, the Hamiltonian is written
\begin{eqnarray}\label{eq2}
H_e = \frac{\omega_e}{2}\sum_j^{N} \sigma_j^z + g\sum_j^{N}(a_{n_j}\sigma_j^+ +a_{n_j}^{\dag}\sigma_j),
\end{eqnarray}
where $\sigma_j^z=|e\rangle_j\langle e|-|g\rangle_j\langle g|$ and $\sigma_j^+=(\sigma_j)^\dag=|e\rangle_j\langle g|$ are the Pauli operators of the $j$th NV center, $g$ is the coupling strength, $n_j$ is the position of the NAMR coupled with $j$th NV center, and $N$ is the number of NV centers.

For periodic boundary conditions (PBC), we can introduce the operators $a_k=\frac{1}{\sqrt{N_r}}\sum_n e^{ikn}a_{n}$, where $k\in[-\pi,\pi]$ is the quasi-momentum of phonons.
In the single-excitation subspace, the nonlinear term in the Hamiltonian~(\ref{eq1}) vanishes. Then,
the nonlinear waveguide is reduced to a coupled-cavity array, and the Hamiltonian can be diagonalized as $H_r=\sum_k\omega_k a_k^\dag a_k$ \cite{PhysRevA.93.033833}.
This subspace only contains single-phonon scattering states with the corresponding eigenenergy
\begin{eqnarray}\label{eq3}
\omega_k^{(1)}=\omega_r-2J\cos(k).
\end{eqnarray}

\begin{figure}
	\includegraphics[width=8cm]{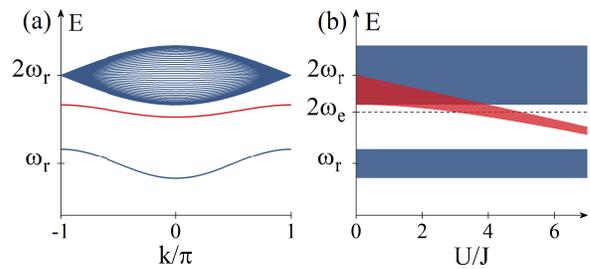}
	\caption{\label{fig2}(Color online)
		(a) Dispersion relation in the one- and two-phonon subspaces with $U=4J$.
		(b) The band structure as a function of the interaction strength $U$. The scattering continuums (blue) and bound-state band (red).
	}
\end{figure}

On the contrary, two-excitation subspace includes scattering states and bound states.
(i) When the phonons are located in different sites, the nonlinear term does not work.
Two free phonons constitute a two-phonon scattering state, whose eigenenergy can be obtained by summing the energy of two single-phonon scattering states \cite{Valiente_2008}
\begin{eqnarray}\label{eq4}
\omega_K^{(2)}&=&\omega_{k_1}^{(1)} +\omega_{k_2}^{(1)} \notag\\
&=&2\omega_r-4J\cos(K/2)\cos(q),
\end{eqnarray}
where $K=k_1+k_2$ and $q=(k_1-k_2)/2$ are the centre-of-mass momentum and the relative momentum of two phonons $k_1,\,k_2$.
(ii) The on-site attractive interaction can bind two phonons into a dimer, i.e., two-phonon bound state.
The dimer can propagate along with the waveguide, tunneling together in adjacent sites, but each phonon cannot move alone.
The bound state with the centre-of-mass momentum $K$ is represented as \cite{PhysRevLett.124.213601}
\begin{eqnarray}\label{eq5}
|\Psi_K\rangle = B_K^\dag|vac\rangle = \frac{1}{\sqrt{2}}\sum_{n,m}\psi_K(n,m)a_n^\dag a_m^\dag|vac\rangle,
\end{eqnarray}
where $B_K^\dag$ is the creation operator of the bound state and the wave function is $\psi_K(n,m)=e^{iK\cdot(n+m)/2}\varphi_K(n-m)/\sqrt{N_c}$.
By solving the stationary Schr\"{o}dinger equation, the corresponding eigenenergy can be obtained
\begin{eqnarray}\label{eq6}
E_K=2\omega_r-\sqrt{U^2+16J^2\cos^2(K/2)}.
\end{eqnarray}
Here the wave function $\varphi_K(n-m)\propto e^{-|n-m|/\lambda_K}$ is exponentially localized around $n-m=0$ and $\lambda_K=1/\text{arcsinh}{(U/4J\cos(K/2))}$ is the width of the wave packet.

We show the dispersion relations and the band structure of the phononic waveguide in Fig.~\ref{fig2}.
This waveguide contains two scattering continuums with finite
bandwidth $4J$ and $8J$, respectively.
As $U$ increases, one band of the bound states gradually separates from the scattering continuums.
When the bound-state energy is below the scattering continuum, the dissociation of the dimer is suppressed since free-particle states with the same energy as the bound state are forbidden in the waveguide.

\section{single- and two NV centers decay}
Previous work has discussed the dissipation dynamics of atoms in the band and band gap of waveguides \cite{PhysRevLett.64.2418,PhysRevB.43.12772,Goban2014,Lambropoulos_2000}.
When the atomic energy is within the scattering continuum, the waveguide will mediate coherent dipole-dipole interactions and collective dissipations, which can be reduced to pure Dicke superradiance under certain conditions \cite{PhysRevLett.110.080502}.
When the atomic frequency is located in the band gap, a long-lived atom-photon dressed state is produced.
Since there is a nonvanishing photon loss rate $\kappa$, the photon mode component of the dressed state induces small atomic decay $\Gamma_1\propto\kappa$ \cite{PhysRevA.93.033833}.

Unlike regular waveguides, there is an extra bound-state band in our scheme.
Different band structure induced by nonlinearity produces novel dissipation dynamics.
The frequency of NV centers is assumed to be below the scattering continuum but within the bound-state band.
It has been shown in  Ref. \cite{PhysRevLett.124.213601} that within the bound-state band a two-photon decay independent of $\kappa$ will be induced, resulting in faster collective radiance than superradiance.

We first focus on the case of two NV centers coupled to the waveguide at the site $n_1$ and $n_2$ under the weak-coupling limit $g\ll J,\,U$.
The single-phonon detuning is $\delta_k=\omega_k-\omega_e$, and $\Delta_K=E_K-2\omega_e$ is the detuning between the two NV centers and the bound state $|\Psi_K\rangle$.
In the rotating frame with $U=e^{-i(\frac{1}{2}\omega_e\sum_j\sigma_j^z + \omega_e\sum_n a_n^\dag a_n)t}$, a general wave function of two-excitation state can be written in the form
\begin{eqnarray}\label{eq7}
|\phi(t)\rangle&=& \Big\{c_e(t)\sigma_1^+\sigma_2^++ \sum_k [c_{1k}(t)\sigma_1^+ + c_{2k}(t)\sigma_2^+]a_k^\dag \notag\\
&& +\sum_K c_K(t)B_K^\dag \Big\}|g,g,vac\rangle,
\end{eqnarray}
where $c_e,c_{ik}$ and $c_K$ are the corresponding probability amplitudes.
Initially, the two NV centers and the waveguide are in the excited state $|e\rangle$ and the vacuum state $|vac\rangle$, respectively.

\begin{figure}
	\includegraphics[width=8cm]{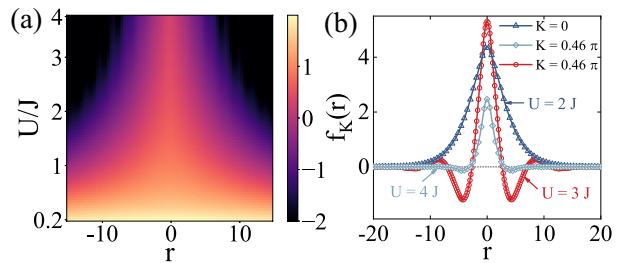}
	\caption{\label{fig3}(Color online)
		(a) $f_{K=0}(n_1,n_2)$ as a function of $U/J$ and $r=n_1-n_2$.
		(b) $f_K(r)$ as a function of $r$ with different $U$ and $K$.
	}
\end{figure}

Since the two-phonon scattering states are off-resonant, they have been neglected and therefore do not appear in  equation~(\ref{eq7}).
By substituting $|\phi(t)\rangle$ into the Schr\"{o}dinger equation, we can obtain the evolution equation of amplitudes.
In the case of $\delta_k\gg|\Delta_K|,g$ and $\Delta_K\approx0$, it allows us to adiabatically eliminate $c_{ik}$, resulting in coupled equations containing only $c_e$ and $c_K$
\begin{eqnarray}\label{eq8}
i\dot{c}_e=-\frac{g^2}{J\sqrt{N_r}}\sum_K e^{iK(n_1+n_2)/2}f_K(n_1,n_2)c_K,
\end{eqnarray}
\begin{eqnarray}\label{eq9}
i\dot{c}_K=\Delta_Kc_K-\frac{g^2}{J\sqrt{N_r}}e^{-iK(n_1+n_2)/2}f_K(n_1,n_2)c_e.
\end{eqnarray}
Here $f_K(n_1,n_2)$ denotes two-phonon correlations and can be calculated by
\begin{eqnarray}\label{eq10}
f_K(n_1,n_2)&=\frac{2\sqrt{2}}{N_r}\sum_k \Big\{ \frac{J\cos[(k-K/2)(n_1-n_2)]}{\delta_k} \notag\\
&\times \frac{\sinh(\lambda_K^{-1})}{[\cosh(\lambda_K^{-1})-\cos(k-K/2)]}\varphi_K(0)\Big\}.
\end{eqnarray}

In Fig.~\ref{fig3}, we plot the correlation $f_K(r)$ versus $U,K$ and $r=n_1-n_2$.
$f_K(r)$ is an even function of $r$ and it's localized around $r=0$.
The correlation is significant only when the two phonons are very close together, while the correlation at a long distance is negligible.
When $K=0$, $f_K(r)$ decreases almost monotonically with $U$ and $|r|$.
When $K\neq0$, the correlation goes back and forth across the zero value, meaning that there is no correlation between the two phonons at the zero positions.

\begin{figure}
	\includegraphics[width=6cm]{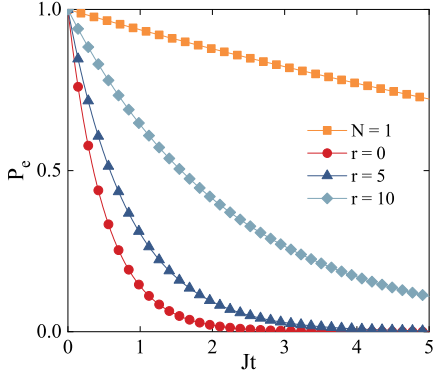}
	\caption{\label{fig4}(Color online)
		Evolution of the excited state population $P_e$ for $N=1$ (orange curves) and $N=2$ (others) NV centers with $\omega_e-\omega_r=-2.03J$, $g=0.1J$, $U=0.7J$ and $\kappa=0.2J$.
		The distance between two NV centers are $r=0,5,10$.
	}
\end{figure}

Under the Markov approximation, we can solve the coupled equations~(\ref{eq8}-\ref{eq9}) and get $P_e(t)=|c_e(t)|^2=|c_e(0)|^2e^{-\Gamma t}$, i.e., the NV centers decay at a rate
\begin{eqnarray}\label{eq11}
\Gamma=\frac{2g^4}{J^2v_g(K_0)}f_{K_0}^2(n_1,n_2),
\end{eqnarray}
where $v_g(K)=4J^2\sin(K)/\sqrt{U^2+16J^2\cos^2(K/2)}$ is the group velocity.
$K_0$ denotes the centre-of-mass momentum of state $|\Psi_{K_0}\rangle$  and satisfies $2\omega_e=E_{K=K_0}$.
It is clear that the dynamics of two close NV centers is dominated by the two-phonon decay as shown in Fig.~\ref{fig4}.
Therefore, this allows us to ignore the single phonon decay  and focus only on the two-phonon process in the following discussion.

\section{collective radiance}
\begin{figure}
	\includegraphics[width=8cm]{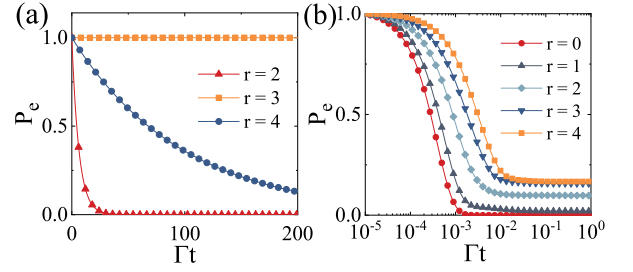}
	\caption{\label{fig5}(Color online)
		(a) Evolution of the excited state population $P_e$ for $N=2$ spins with $K_0=0.46\pi$ and $U=4$.
		The curve at $r=2,4$ decays exponentially, while the spins at spacing $r=3$ stay in the excited state for a long time.
		(b) Evolution of the excited state population for four NV centers arranged at equal spacing.
		The parameters are $\omega_e-\omega_r=-2.04J$, $g=0.1J$, $U=J$ and distance $r=0,1,2,3,4$.
	}
\end{figure}
From equations~(\ref{eq8}-\ref{eq9}), we can write the effective Hamiltonian $H_{eff}$ of the system
\begin{eqnarray}\label{eq12}
H_{eff} = \sum_K \Delta_K B_K^\dag B_K + \frac{g^2}{J} \sum_{i,j}^{N}[\sigma_i^-\sigma_j^-B_{n_i,n_j}^\dag+H.c.],  \notag\\
\end{eqnarray}
where $B_{n_i,n_j}^\dag=\frac{1}{\sqrt{N}}\sum_K e^{-iK(n_i+n_j)/2}f_K(n_i,n_j)B_K^\dag$ is the creation operator of  the two-phonon bound state at the position $n_i$ and $n_j$.
Since the nonlinearity reduces the total energy of $n$ phonons in the same position by $n(n-1)U/2$, only one bound state of two phonons is allowed at a position in our frequency range.

In the previous derivation, we used the Markov approximation, which only holds if the group velocity $v_g(K)$ is much greater than the coupling strength $g^2/J$.
As long as the frequency of NV centers is far away from $\frac{1}{2}E_{K=0}$, $v_g(K)\sim J\gg g^2/J$ is satisfied.
Assuming the average phonon number $\langle n\rangle\approx0$ at low temperatures, the Markov master equation can be obtained
\begin{eqnarray}\label{eq13}
\dot{\rho}&=&-i\bigg[\frac{1}{2}\sum_{i,j,k,l}\left(U_{ij,kl}\sigma_k^+\sigma_l^+\sigma_i^-\sigma_j^-+H.c.\right),\, \rho\bigg] \notag\\
&+&\sum_{i,j,k,l}\frac{1}{2}\Gamma_{ij,kl}\big(\sigma_i^-\sigma_j^-\rho\sigma_k^+\sigma_l^+ -\rho\sigma_k^+\sigma_l^+\sigma_i^-\sigma_j^-\big)+H.c., \notag\\
\end{eqnarray}
where $\left[\cdot,\cdot\right]$ denotes the commutator.
The decay rate $\Gamma_{ij,kl}=\Gamma_0\text{Re}\{A_{ij,kl}\}$ and coherent coupling $U_{ij,kl} = \frac{1}{2}\Gamma_0\text{Im}\{A_{ij,kl}\}$ are calculated by
\begin{eqnarray}\label{eq14}
A_{ij,kl} &=& f_{K_0}(n_i-n_j)f_{K_0}(n_k-n_l) \notag\\
& & \times e^{iK_0|(n_k+n_l)-(n_i+n_j)|/2},
\end{eqnarray}
where $\Gamma_0=2g^4/J^2v_g(K_0)$

\subsection{Subradiance}

From the previous section we can see that the two spins decay exponentially with a rate proportional to $f_{K_0}^2(r)$.
However, in Fig.~\ref{fig3}(b), there are several special positions that satisfy $f_{K_0}(r)=0$, indicating the appearance of long-life states.
In Fig.~\ref{fig5}(a), we plot the decay dynamics of two particles at a distance $r=2,3,4$ with $U=4J$ and $K_0\approx0.46\pi$.
The usual exponential decay occurs at $r=2,4$.
But, when $r=3$, the two spins stay in excited state for a long time, and it is a subradiant state of the system.

On the other hand, the existence of long-distance interference in the system, $A_{ij,kl}\propto e^{iK_0|(n_k+n_l)-(n_i+n_j)|/2}$, indicates that there may be accelerated or suppressed radiance dynamics in the multi-particle system.
An example of four-particle subradiant states is shown in Fig.~\ref{fig5}(b).
All NV centers are equally spaced with a distance $r=0,1,2,3,4$.
In the initial period the spins decay rapidly.
However, when $r\neq 0$, $P_e$ decays to a nonvanishing value and stays there for a long time.
The residual excitation comes from the subradiant state $|D_2\rangle$ \cite{PhysRevLett.124.213601}, and other subradiant states contribute very little.
The subradiant states like $|D_2\rangle$ are widely present in the multi-particle system with $r\neq0$.

\begin{figure}
	\includegraphics[width=8cm]{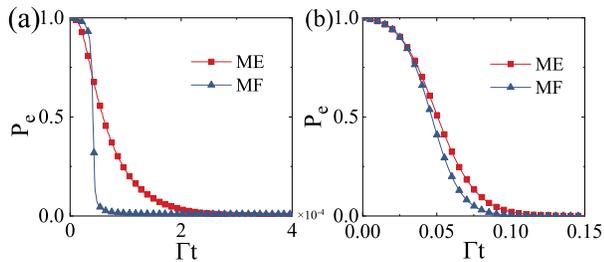}
	\caption{\label{fig6}(Color online)
		Master equation simulation and mean-field approximation for (a) supercorrelated radiance in Eq.~(\ref{eq15}) and (b) superradiance in Eq.~(\ref{eq16}) with $N=100$ spins.
	}
\end{figure}

\subsection{Supercorrelated radiance}
When all NV centers are in the same site, the above subradiant states cannot exist.
Instead, they emit phonons at a rate faster than Dicke superradiance.
Since $A_{ij,kl}=\Gamma_0f_{K_0}^2(0)$ is a real and uniform number, we can introduce the collective operator $S_+=(S_-)^\dag =\sum_j \sigma_j^+$ and get the master equation
\begin{eqnarray}\label{eq15}
\dot{\rho}=\frac{\Gamma}{2}\left(2S_-^2\rho S_+^2 - \rho S_+^2 S_-^2 - S_+^2 S_-^2 \rho \right).
\end{eqnarray}
It is very similar to Dicke superradiance, which is described by the master equation
\begin{eqnarray}\label{eq16}
\dot{\rho}=\frac{\Gamma_D}{2}\left(2S_-\rho S_+ - \rho S_+ S_- - S_+ S_- \rho \right).
\end{eqnarray}
Dicke superradiance describes  the effective emission rate $\Gamma_{eff}\propto \langle S_+S_-\rangle$ increases by $N$ to $N^2$ times under the collective effect.
However, the emission rate $\Gamma_{eff}\propto \langle S_+^2S_-^2\rangle$ described in Equation~(\ref{eq15}) increases by a factor of $N^2$ to $N^4$, which is also known as the supercorrelated radiance \cite{PhysRevLett.124.213601}.
In Fig.~\ref{fig6}, we compare the dynamic evolution of the supercorrelated radiance and Dicke superradiance by using the master equation and the mean-field approximation.
In supercorrelated radiance, the spins decay more rapidly, and the lifetime is reduced by $N$ times compared with the superradiance.
The mean-field approximation is valid for superradiance, but it fails in the description of supercorrelated radiance.
The deviation suggests that supercorrelated radiance is very different from superradiance.

\begin{figure}
	\includegraphics[width=8cm]{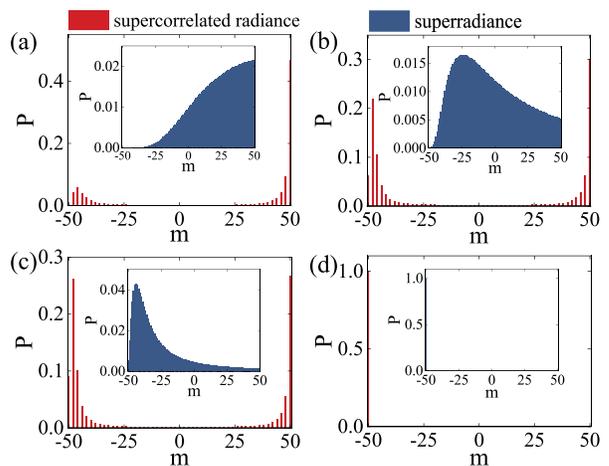}
	\caption{\label{fig7}(Color online)
		(a)-(d) The temporal distribution of states at four moments in evolution for supercorrelated radiance.
		The results of superradiance is shown in inset.
	}
\end{figure}

In order to gain more insight into the supercorrelated radiance, we show the state distribution at four moments in the evolution of the master equation, where $m$ represents Dicke states $|S =N/2,m=S_z\rangle$ in Fig.~\ref{fig7}.
Assuming that the initial state is fully excited, the state at this moment must only be at $|m=50\rangle$, so it is not shown in the figure.
For Dicke surperradiance, the peak of the state distribution moves continuously from $|m=50\rangle$ to $|m=-50\rangle$ over time. Surprisingly, the state in supercorrelated radiance is approximately jumping directly from $|m=50\rangle$ to $|m=- 50\rangle$.
These intermediate states have such ultrafast decay rates that they are like a bridge to connect the states $|m=\pm 50\rangle$ but nearly unoccupied.
Finally, All distributions go to zero except for $m=-50$, which means all NV centers go back to the ground state $|g\rangle$.
In addition, since the dynamics in supercorrelated radiance consists of only two-photon processes, the population of all odd m is always 0.

\subsection{Reproducing superradiance}
The relationship between supercorrelated radiance and superradiance may not stop there.
The increase in their emission rate both can be attributed to long-distance coherence.
We know that in superradiance one atom is an emitter, and in supercorrelated radiance, pairs of atoms decay collectively.
Arbitrariness in the selection of emission primitives will result in a factor that increases with the number of particles $N$, and coincidentally, the decay rate in supercorrelated radiance is about $N$ times the average decay rate of superradiance.
It inspires us that if we cancel the arbitrariness when emitting phonons, it is possible to reproduce Dicke superradiance.

\begin{figure}
	\includegraphics[width=8cm]{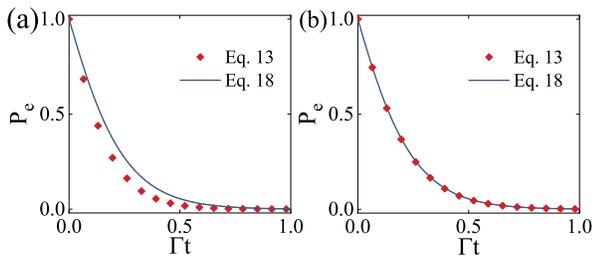}
	\caption{\label{fig8}(Color online)
		Compare (a) $U=3J$ and (b) $U=4J$ by the results of Eq.~(\ref{eq13}) and Eq.~(\ref{eq18}).
		The NV centers are placed at $[0,0,4,4]$.
		The parameters are $K_0=\frac{1}{2}\pi$ and $\Gamma_D=4\Gamma$.
	}
\end{figure}

As an example, we place four NV centers with frequency $\omega_e=\frac{1}{2}E_{K_0=0.5\pi}$ at sites $n_1=n_2=0$ and $n_3=n_4=4$.
In this case, the two-phonon correlation $f_{K_0}(r)$ has only two possible values, $f_{K_0}(0)$ and $f_{K_0}(4)$.
If $f_{K_0}(4)\approx 0\ll f_{K_0}(0)$ is satisfied, NV center 1 can only cooperate with NV center 2 to emit phonon pairs, but not with NV centers 3 or 4, and vice versa.
So the NV centers are divided into two groups $\alpha=\{1,2\}$ and $\beta=\{3,4\}$.
The two-phonon correlation occurs only within each group, not between groups.
This limits the possible emission primitives to the NV centers within each group, while others are prohibited.
By ignoring all the terms in equation~(\ref{eq13}) that contain $f_{K_0}(4)$ and introducing group annihilation operators $\sigma_\alpha=\frac{1}{2}(\sigma_1^-\sigma_2^-+\sigma_2^-\sigma_1^-),\sigma_\beta=\frac{1}{2}(\sigma_3^-\sigma_4^-+\sigma_4^-\sigma_3^-)$, we get
\begin{eqnarray}\label{eq17}
\dot{\rho}= \sum_{\mu,\nu}2\Gamma \left(\sigma_\mu\rho\sigma_\nu^\dag -\sigma_\nu^\dag\sigma_\mu\rho\right) + H.c.,
\end{eqnarray}
where $\mu,\nu\in \{\alpha,\beta\}$ denote the group number and $\Gamma=A_{ij,kl}=\Gamma_0f_{K_0}^2(0)$.
If we further introduce the corresponding collective operators $S_\Lambda=\sum_\mu\sigma_\mu$, the equation~(\ref{eq17}) is reduced to
\begin{eqnarray}\label{eq18}
\dot{\rho}=2\Gamma\left(2S_\Lambda\rho S_\Lambda^\dag - \rho S_\Lambda^\dag S_\Lambda - S_\Lambda^\dag S_\Lambda \rho \right).
\end{eqnarray}
This is completely consistent with the master equation of superradiance with $N/2$ particles and the decay rate $\Gamma_D=4\Gamma$.
In Fig.~\ref{fig8}, we compare the results of equation~(\ref{eq13}) and equation~(\ref{eq18}) with $U=3J$ and $U=4J$.
When $U=4J$, the curve fits perfectly, indicating that the reproducing Dicke superradiance is successful.
When $U=3J$, the curve deviates because $f_{K_0}(4)\ll f_{K_0}(0)$ is not fully satisfied.

In fact, as long as an even number of NV centers are grouped in pairs, with sufficient spacing between each group, the two-phonon correlation between groups can be cancelled.
In order to eliminate the coherent coupling term in equation~(\ref{eq13}), a feasible solution is to take $K_0=2\pi/a$ and the spacing is an integer multiple of $a$.
In this case, a group has only two possible energy levels to occupy, all in the excited state and all in the ground state.
There is one-to-one correspondence between the system and the superradiance: mapping $N/2$ groups to $N/2$ atoms, the bound-state band to the scattering continuum, and the two-phonon interaction to the JC model.
Thus, the system naturally inherits the properties of superradiance, where the factor $4$ in the radiance rate is derived from the exchange symmetry of NV centers within the group.
Therefore, the essence of supercorrelated radiance is still interference.

\section{EXPERIMENTAL FEASIBILITY }
In order to examine the feasibility of this scheme in experiment, we now discuss the relevant parameters.
For the mechanical resonator, we consider a single-crystal diamond beam with dimensions $(L,w,h)=(850,80,80)$ $\text{nm}$.
The material properties are Young's modulus $E=1200$ GPa and mass density $\varrho\approx3500$ kg $\text{m}^{-3}$.
The fundamental frequency for the bending mode and the zero-point motion are $\omega_r =4.73^2 \times (h/L^2)\sqrt{E/12\varrho}\approx2\pi\times2$ GHz and $z_0=\sqrt{\hbar/2m\omega_r}$.
Since the single-crystal diamond nanomechanical resonators
with quality factors $Q$ exceeding one million have been realized experimentally, we can obtain the resonators decay rate $\kappa=\omega_r/Q\sim2$ kHz.
Each resonator is charged and interacts capacitatively with two nearby wires.
The two capacitors $C(z_n)$ are dependent on the position $z_n=z_0(a_n+a_n^\dag)$, one of which introduces the interaction between different resonators, and the other couples the resonator with an auxiliary superconducting qubit.
The coupling between nearest-neighbor NAMRs can be as large as 1 MHz when the applied voltage reaches 10 V \cite{Rabl2010}.
In the case of large detuning between the NAMR and the superconducting qubit, the superconducting qubit with a strong driving field will be in the dressed ground state $|-\rangle$ all the time and introduces nonlinearity up to 8 MHz \cite{PhysRevA.82.032101}.

By placing the NV centers near the surface of the center of long axis of the beam, the strain coupling strength between a single NV spin and the resonator can reach the maximum $g/2\pi\sim 1$ kHz \cite{PhysRevLett.110.156402}.
The coupling between nearest-neighbor NAMR $J/2\pi=10$ kHz and the nonlinear strength only need tens of kHz, which is easy to achieve.
Since the weak coupling limit $g\ll J$ is satisfied, this allows us to use the Born-Markov approximation.
Assuming that the entire device is in a diluting refrigerator with temperature $T\sim10$ mK, the average thermal phonon number is $\bar{n}=1/\left(e^{\hbar \omega_r/k_BT}-1\right)\sim0$.
At low temperature, the spin relaxation time $T_1$ of NV centers is very long, up to several seconds, so the decoherence mainly comes from the spin dephasing $T_2$.
The evolution time is $t=5/J=500$ $\mu\text{s}$ in Fig.~\ref{fig4}, which is shorter than the spin dephasing time $T_2$ \cite{Bar-Gill2013,Balasubramanian2009,Maurer1283}.
And the time decreases rapidly with the increase of the NV center number $N$.

\section{Conclusion}
In conclusion, we have proposed a scheme to explore the collective radiance effect in a nonlinear phononic waveguide.
We find a  bound state and two-phonon decay channel, which can lead to suprcorrelated radiance.
More importantly, the suprcorrelated radiance exhibits a much faster emission rate than the superradiance, and the peak of state distribution approximatively jumps directly from $|m=N/2\rangle$ to $|m=-N/2\rangle$.
In addition, we discuss how to reproduce superradiance in this nonlinear environment, which shows that the essence of supercorrelated radiance is still interference.

\section*{Acknowledgments}

This work is
supported by the National Natural Science Foundation of
China under Grant No. 92065105  and
Natural Science Basic Research Program of Shaanxi
(Program No. 2020JC-02).

\begin{appendix}

\end{appendix}
%\bibliography{ref}
%\bibliography{ref}
%\bibliography{ref}

\begin{thebibliography}{91}%
\makeatletter
\providecommand \@ifxundefined [1]{%
 \@ifx{#1\undefined}
}%
\providecommand \@ifnum [1]{%
 \ifnum #1\expandafter \@firstoftwo
 \else \expandafter \@secondoftwo
 \fi
}%
\providecommand \@ifx [1]{%
 \ifx #1\expandafter \@firstoftwo
 \else \expandafter \@secondoftwo
 \fi
}%
\providecommand \natexlab [1]{#1}%
\providecommand \enquote  [1]{``#1''}%
\providecommand \bibnamefont  [1]{#1}%
\providecommand \bibfnamefont [1]{#1}%
\providecommand \citenamefont [1]{#1}%
\providecommand \href@noop [0]{\@secondoftwo}%
\providecommand \href [0]{\begingroup \@sanitize@url \@href}%
\providecommand \@href[1]{\@@startlink{#1}\@@href}%
\providecommand \@@href[1]{\endgroup#1\@@endlink}%
\providecommand \@sanitize@url [0]{\catcode `\\12\catcode `\$12\catcode
  `\&12\catcode `\#12\catcode `\^12\catcode `\_12\catcode `\%12\relax}%
\providecommand \@@startlink[1]{}%
\providecommand \@@endlink[0]{}%
\providecommand \url  [0]{\begingroup\@sanitize@url \@url }%
\providecommand \@url [1]{\endgroup\@href {#1}{\urlprefix }}%
\providecommand \urlprefix  [0]{URL }%
\providecommand \Eprint [0]{\href }%
\providecommand \doibase [0]{https://doi.org/}%
\providecommand \selectlanguage [0]{\@gobble}%
\providecommand \bibinfo  [0]{\@secondoftwo}%
\providecommand \bibfield  [0]{\@secondoftwo}%
\providecommand \translation [1]{[#1]}%
\providecommand \BibitemOpen [0]{}%
\providecommand \bibitemStop [0]{}%
\providecommand \bibitemNoStop [0]{.\EOS\space}%
\providecommand \EOS [0]{\spacefactor3000\relax}%
\providecommand \BibitemShut  [1]{\csname bibitem#1\endcsname}%
\let\auto@bib@innerbib\@empty
%</preamble>
\bibitem [{\citenamefont {Dicke}(1954)}]{PhysRev.93.99}%
  \BibitemOpen
  \bibfield  {author} {\bibinfo {author} {\bibfnamefont {R.~H.}\ \bibnamefont
  {Dicke}},\ }\bibfield  {title} {\bibinfo {title} {Coherence in spontaneous
  radiation processes},\ }\href {https://doi.org/10.1103/PhysRev.93.99}
  {\bibfield  {journal} {\bibinfo  {journal} {Phys. Rev.}\ }\textbf {\bibinfo
  {volume} {93}},\ \bibinfo {pages} {99} (\bibinfo {year} {1954})}\BibitemShut
  {NoStop}%
\bibitem [{\citenamefont {Gross}\ and\ \citenamefont
  {Haroche}(1982)}]{GROSS1982301}%
  \BibitemOpen
  \bibfield  {author} {\bibinfo {author} {\bibfnamefont {M.}~\bibnamefont
  {Gross}}\ and\ \bibinfo {author} {\bibfnamefont {S.}~\bibnamefont
  {Haroche}},\ }\bibfield  {title} {\bibinfo {title} {Superradiance: An essay
  on the theory of collective spontaneous emission},\ }\href
  {https://doi.org/https://doi.org/10.1016/0370-1573(82)90102-8} {\bibfield
  {journal} {\bibinfo  {journal} {Phys. Rep.}\ }\textbf {\bibinfo {volume}
  {93}},\ \bibinfo {pages} {301} (\bibinfo {year} {1982})}\BibitemShut
  {NoStop}%
\bibitem [{\citenamefont {Rehler}\ and\ \citenamefont
  {Eberly}(1971)}]{PhysRevA.3.1735}%
  \BibitemOpen
  \bibfield  {author} {\bibinfo {author} {\bibfnamefont {N.~E.}\ \bibnamefont
  {Rehler}}\ and\ \bibinfo {author} {\bibfnamefont {J.~H.}\ \bibnamefont
  {Eberly}},\ }\bibfield  {title} {\bibinfo {title} {Superradiance},\ }\href
  {https://doi.org/10.1103/PhysRevA.3.1735} {\bibfield  {journal} {\bibinfo
  {journal} {Phys. Rev. A}\ }\textbf {\bibinfo {volume} {3}},\ \bibinfo {pages}
  {1735} (\bibinfo {year} {1971})}\BibitemShut {NoStop}%
\bibitem [{\citenamefont {Kirton}\ and\ \citenamefont
  {Keeling}(2017)}]{PhysRevLett.118.123602}%
  \BibitemOpen
  \bibfield  {author} {\bibinfo {author} {\bibfnamefont {P.}~\bibnamefont
  {Kirton}}\ and\ \bibinfo {author} {\bibfnamefont {J.}~\bibnamefont
  {Keeling}},\ }\bibfield  {title} {\bibinfo {title} {Suppressing and restoring
  the dicke superradiance transition by dephasing and decay},\ }\href
  {https://doi.org/10.1103/PhysRevLett.118.123602} {\bibfield  {journal}
  {\bibinfo  {journal} {Phys. Rev. Lett.}\ }\textbf {\bibinfo {volume} {118}},\
  \bibinfo {pages} {123602} (\bibinfo {year} {2017})}\BibitemShut {NoStop}%
\bibitem [{\citenamefont {Pavolini}\ \emph {et~al.}(1985)\citenamefont
  {Pavolini}, \citenamefont {Crubellier}, \citenamefont {Pillet}, \citenamefont
  {Cabaret},\ and\ \citenamefont {Liberman}}]{PhysRevLett.54.1917}%
  \BibitemOpen
  \bibfield  {author} {\bibinfo {author} {\bibfnamefont {D.}~\bibnamefont
  {Pavolini}}, \bibinfo {author} {\bibfnamefont {A.}~\bibnamefont
  {Crubellier}}, \bibinfo {author} {\bibfnamefont {P.}~\bibnamefont {Pillet}},
  \bibinfo {author} {\bibfnamefont {L.}~\bibnamefont {Cabaret}},\ and\ \bibinfo
  {author} {\bibfnamefont {S.}~\bibnamefont {Liberman}},\ }\bibfield  {title}
  {\bibinfo {title} {Experimental evidence for subradiance},\ }\href
  {https://doi.org/10.1103/PhysRevLett.54.1917} {\bibfield  {journal} {\bibinfo
   {journal} {Phys. Rev. Lett.}\ }\textbf {\bibinfo {volume} {54}},\ \bibinfo
  {pages} {1917} (\bibinfo {year} {1985})}\BibitemShut {NoStop}%
\bibitem [{\citenamefont {DeVoe}\ and\ \citenamefont
  {Brewer}(1996)}]{PhysRevLett.76.2049}%
  \BibitemOpen
  \bibfield  {author} {\bibinfo {author} {\bibfnamefont {R.~G.}\ \bibnamefont
  {DeVoe}}\ and\ \bibinfo {author} {\bibfnamefont {R.~G.}\ \bibnamefont
  {Brewer}},\ }\bibfield  {title} {\bibinfo {title} {Observation of
  superradiant and subradiant spontaneous emission of two trapped ions},\
  }\href {https://doi.org/10.1103/PhysRevLett.76.2049} {\bibfield  {journal}
  {\bibinfo  {journal} {Phys. Rev. Lett.}\ }\textbf {\bibinfo {volume} {76}},\
  \bibinfo {pages} {2049} (\bibinfo {year} {1996})}\BibitemShut {NoStop}%
\bibitem [{\citenamefont {Guerin}\ \emph {et~al.}(2016)\citenamefont {Guerin},
  \citenamefont {Ara\'ujo},\ and\ \citenamefont
  {Kaiser}}]{PhysRevLett.116.083601}%
  \BibitemOpen
  \bibfield  {author} {\bibinfo {author} {\bibfnamefont {W.}~\bibnamefont
  {Guerin}}, \bibinfo {author} {\bibfnamefont {M.~O.}\ \bibnamefont
  {Ara\'ujo}},\ and\ \bibinfo {author} {\bibfnamefont {R.}~\bibnamefont
  {Kaiser}},\ }\bibfield  {title} {\bibinfo {title} {Subradiance in a large
  cloud of cold atoms},\ }\href
  {https://doi.org/10.1103/PhysRevLett.116.083601} {\bibfield  {journal}
  {\bibinfo  {journal} {Phys. Rev. Lett.}\ }\textbf {\bibinfo {volume} {116}},\
  \bibinfo {pages} {083601} (\bibinfo {year} {2016})}\BibitemShut {NoStop}%
\bibitem [{\citenamefont {Scully}(2015)}]{PhysRevLett.115.243602}%
  \BibitemOpen
  \bibfield  {author} {\bibinfo {author} {\bibfnamefont {M.~O.}\ \bibnamefont
  {Scully}},\ }\bibfield  {title} {\bibinfo {title} {Single photon subradiance:
  Quantum control of spontaneous emission and ultrafast readout},\ }\href
  {https://doi.org/10.1103/PhysRevLett.115.243602} {\bibfield  {journal}
  {\bibinfo  {journal} {Phys. Rev. Lett.}\ }\textbf {\bibinfo {volume} {115}},\
  \bibinfo {pages} {243602} (\bibinfo {year} {2015})}\BibitemShut {NoStop}%
\bibitem [{\citenamefont {Kovalev}\ \emph {et~al.}(2018)\citenamefont
  {Kovalev}, \citenamefont {Wang}, \citenamefont {Deinert}, \citenamefont
  {Awari}, \citenamefont {Chen}, \citenamefont {Green}, \citenamefont
  {Germanskiy}, \citenamefont {de~Oliveira}, \citenamefont {Lee}, \citenamefont
  {Deac}, \citenamefont {Turchinovich}, \citenamefont {Stojanovic},
  \citenamefont {Eisebitt}, \citenamefont {Radu}, \citenamefont {Bonetti},
  \citenamefont {Kampfrath},\ and\ \citenamefont {Gensch}}]{Kovalev_2018}%
  \BibitemOpen
  \bibfield  {author} {\bibinfo {author} {\bibfnamefont {S.}~\bibnamefont
  {Kovalev}}, \bibinfo {author} {\bibfnamefont {Z.}~\bibnamefont {Wang}},
  \bibinfo {author} {\bibfnamefont {J.-C.}\ \bibnamefont {Deinert}}, \bibinfo
  {author} {\bibfnamefont {N.}~\bibnamefont {Awari}}, \bibinfo {author}
  {\bibfnamefont {M.}~\bibnamefont {Chen}}, \bibinfo {author} {\bibfnamefont
  {B.}~\bibnamefont {Green}}, \bibinfo {author} {\bibfnamefont
  {S.}~\bibnamefont {Germanskiy}}, \bibinfo {author} {\bibfnamefont {T.~V.
  A.~G.}\ \bibnamefont {de~Oliveira}}, \bibinfo {author} {\bibfnamefont
  {J.~S.}\ \bibnamefont {Lee}}, \bibinfo {author} {\bibfnamefont
  {A.}~\bibnamefont {Deac}}, \bibinfo {author} {\bibfnamefont {D.}~\bibnamefont
  {Turchinovich}}, \bibinfo {author} {\bibfnamefont {N.}~\bibnamefont
  {Stojanovic}}, \bibinfo {author} {\bibfnamefont {S.}~\bibnamefont
  {Eisebitt}}, \bibinfo {author} {\bibfnamefont {I.}~\bibnamefont {Radu}},
  \bibinfo {author} {\bibfnamefont {S.}~\bibnamefont {Bonetti}}, \bibinfo
  {author} {\bibfnamefont {T.}~\bibnamefont {Kampfrath}},\ and\ \bibinfo
  {author} {\bibfnamefont {M.}~\bibnamefont {Gensch}},\ }\bibfield  {title}
  {\bibinfo {title} {Selective {THz} control of magnetic order: new
  opportunities from superradiant undulator sources},\ }\href
  {https://doi.org/10.1088/1361-6463/aaac75} {\bibfield  {journal} {\bibinfo
  {journal} {J. Phys. D: Appl. Phys.}\ }\textbf {\bibinfo {volume} {51}},\
  \bibinfo {pages} {114007} (\bibinfo {year} {2018})}\BibitemShut {NoStop}%
\bibitem [{\citenamefont {Bin}\ \emph {et~al.}(2021)\citenamefont {Bin},
  \citenamefont {Wu},\ and\ \citenamefont {L{\"u}}}]{bin2021parity}%
  \BibitemOpen
  \bibfield  {author} {\bibinfo {author} {\bibfnamefont {Q.}~\bibnamefont
  {Bin}}, \bibinfo {author} {\bibfnamefont {Y.}~\bibnamefont {Wu}},\ and\
  \bibinfo {author} {\bibfnamefont {X.-Y.}\ \bibnamefont {L{\"u}}},\ }\bibfield
   {title} {\bibinfo {title} {Parity-symmetry-protected multiphoton bundle
  emission},\ }\href {https://doi.org/0} {\bibfield  {journal} {\bibinfo
  {journal} {Phys. Rev. Lett.}\ }\textbf {\bibinfo {volume} {127}},\ \bibinfo
  {pages} {073602} (\bibinfo {year} {2021})}\BibitemShut {NoStop}%
\bibitem [{\citenamefont {Asenjo-Garcia}\ \emph {et~al.}(2017)\citenamefont
  {Asenjo-Garcia}, \citenamefont {Moreno-Cardoner}, \citenamefont {Albrecht},
  \citenamefont {Kimble},\ and\ \citenamefont {Chang}}]{PhysRevX.7.031024}%
  \BibitemOpen
  \bibfield  {author} {\bibinfo {author} {\bibfnamefont {A.}~\bibnamefont
  {Asenjo-Garcia}}, \bibinfo {author} {\bibfnamefont {M.}~\bibnamefont
  {Moreno-Cardoner}}, \bibinfo {author} {\bibfnamefont {A.}~\bibnamefont
  {Albrecht}}, \bibinfo {author} {\bibfnamefont {H.~J.}\ \bibnamefont
  {Kimble}},\ and\ \bibinfo {author} {\bibfnamefont {D.~E.}\ \bibnamefont
  {Chang}},\ }\bibfield  {title} {\bibinfo {title} {Exponential improvement in
  photon storage fidelities using subradiance and ``selective radiance" in
  atomic arrays},\ }\href {https://doi.org/10.1103/PhysRevX.7.031024}
  {\bibfield  {journal} {\bibinfo  {journal} {Phys. Rev. X}\ }\textbf {\bibinfo
  {volume} {7}},\ \bibinfo {pages} {031024} (\bibinfo {year}
  {2017})}\BibitemShut {NoStop}%
\bibitem [{\citenamefont {Cipris}\ \emph {et~al.}(2021)\citenamefont {Cipris},
  \citenamefont {Moreira}, \citenamefont {do~Espirito~Santo}, \citenamefont
  {Weiss}, \citenamefont {Villas-Boas}, \citenamefont {Kaiser}, \citenamefont
  {Guerin},\ and\ \citenamefont {Bachelard}}]{PhysRevLett.126.103604}%
  \BibitemOpen
  \bibfield  {author} {\bibinfo {author} {\bibfnamefont {A.}~\bibnamefont
  {Cipris}}, \bibinfo {author} {\bibfnamefont {N.~A.}\ \bibnamefont {Moreira}},
  \bibinfo {author} {\bibfnamefont {T.~S.}\ \bibnamefont {do~Espirito~Santo}},
  \bibinfo {author} {\bibfnamefont {P.}~\bibnamefont {Weiss}}, \bibinfo
  {author} {\bibfnamefont {C.~J.}\ \bibnamefont {Villas-Boas}}, \bibinfo
  {author} {\bibfnamefont {R.}~\bibnamefont {Kaiser}}, \bibinfo {author}
  {\bibfnamefont {W.}~\bibnamefont {Guerin}},\ and\ \bibinfo {author}
  {\bibfnamefont {R.}~\bibnamefont {Bachelard}},\ }\bibfield  {title} {\bibinfo
  {title} {Subradiance with saturated atoms: Population enhancement of the
  long-lived states},\ }\href {https://doi.org/10.1103/PhysRevLett.126.103604}
  {\bibfield  {journal} {\bibinfo  {journal} {Phys. Rev. Lett.}\ }\textbf
  {\bibinfo {volume} {126}},\ \bibinfo {pages} {103604} (\bibinfo {year}
  {2021})}\BibitemShut {NoStop}%
\bibitem [{\citenamefont {Guimond}\ \emph {et~al.}(2019)\citenamefont
  {Guimond}, \citenamefont {Grankin}, \citenamefont {Vasilyev}, \citenamefont
  {Vermersch},\ and\ \citenamefont {Zoller}}]{PhysRevLett.122.093601}%
  \BibitemOpen
  \bibfield  {author} {\bibinfo {author} {\bibfnamefont {P.-O.}\ \bibnamefont
  {Guimond}}, \bibinfo {author} {\bibfnamefont {A.}~\bibnamefont {Grankin}},
  \bibinfo {author} {\bibfnamefont {D.~V.}\ \bibnamefont {Vasilyev}}, \bibinfo
  {author} {\bibfnamefont {B.}~\bibnamefont {Vermersch}},\ and\ \bibinfo
  {author} {\bibfnamefont {P.}~\bibnamefont {Zoller}},\ }\bibfield  {title}
  {\bibinfo {title} {Subradiant bell states in distant atomic arrays},\ }\href
  {https://doi.org/10.1103/PhysRevLett.122.093601} {\bibfield  {journal}
  {\bibinfo  {journal} {Phys. Rev. Lett.}\ }\textbf {\bibinfo {volume} {122}},\
  \bibinfo {pages} {093601} (\bibinfo {year} {2019})}\BibitemShut {NoStop}%
\bibitem [{\citenamefont {Facchinetti}\ \emph {et~al.}(2016)\citenamefont
  {Facchinetti}, \citenamefont {Jenkins},\ and\ \citenamefont
  {Ruostekoski}}]{PhysRevLett.117.243601}%
  \BibitemOpen
  \bibfield  {author} {\bibinfo {author} {\bibfnamefont {G.}~\bibnamefont
  {Facchinetti}}, \bibinfo {author} {\bibfnamefont {S.~D.}\ \bibnamefont
  {Jenkins}},\ and\ \bibinfo {author} {\bibfnamefont {J.}~\bibnamefont
  {Ruostekoski}},\ }\bibfield  {title} {\bibinfo {title} {Storing light with
  subradiant correlations in arrays of atoms},\ }\href
  {https://doi.org/10.1103/PhysRevLett.117.243601} {\bibfield  {journal}
  {\bibinfo  {journal} {Phys. Rev. Lett.}\ }\textbf {\bibinfo {volume} {117}},\
  \bibinfo {pages} {243601} (\bibinfo {year} {2016})}\BibitemShut {NoStop}%
\bibitem [{\citenamefont {Ballantine}\ and\ \citenamefont
  {Ruostekoski}(2020)}]{PhysRevResearch.2.023086}%
  \BibitemOpen
  \bibfield  {author} {\bibinfo {author} {\bibfnamefont {K.~E.}\ \bibnamefont
  {Ballantine}}\ and\ \bibinfo {author} {\bibfnamefont {J.}~\bibnamefont
  {Ruostekoski}},\ }\bibfield  {title} {\bibinfo {title} {Subradiance-protected
  excitation spreading in the generation of collimated photon emission from an
  atomic array},\ }\href {https://doi.org/10.1103/PhysRevResearch.2.023086}
  {\bibfield  {journal} {\bibinfo  {journal} {Phys. Rev. Research}\ }\textbf
  {\bibinfo {volume} {2}},\ \bibinfo {pages} {023086} (\bibinfo {year}
  {2020})}\BibitemShut {NoStop}%
\bibitem [{\citenamefont {Arcizet}\ \emph {et~al.}(2006)\citenamefont
  {Arcizet}, \citenamefont {Cohadon}, \citenamefont {Briant}, \citenamefont
  {Pinard},\ and\ \citenamefont {Heidmann}}]{Arcizet2006}%
  \BibitemOpen
  \bibfield  {author} {\bibinfo {author} {\bibfnamefont {O.}~\bibnamefont
  {Arcizet}}, \bibinfo {author} {\bibfnamefont {P.-F.}\ \bibnamefont
  {Cohadon}}, \bibinfo {author} {\bibfnamefont {T.}~\bibnamefont {Briant}},
  \bibinfo {author} {\bibfnamefont {M.}~\bibnamefont {Pinard}},\ and\ \bibinfo
  {author} {\bibfnamefont {A.}~\bibnamefont {Heidmann}},\ }\bibfield  {title}
  {\bibinfo {title} {Radiation-pressure cooling and optomechanical instability
  of a micromirror},\ }\href {https://doi.org/10.1038/nature05244} {\bibfield
  {journal} {\bibinfo  {journal} {Nature}\ }\textbf {\bibinfo {volume} {444}},\
  \bibinfo {pages} {71} (\bibinfo {year} {2006})}\BibitemShut {NoStop}%
\bibitem [{\citenamefont {Schliesser}\ \emph {et~al.}(2008)\citenamefont
  {Schliesser}, \citenamefont {Rivi\`{e}re}, \citenamefont {Anetsberger},
  \citenamefont {Arcizet},\ and\ \citenamefont {Kippenberg}}]{Schliesser2008}%
  \BibitemOpen
  \bibfield  {author} {\bibinfo {author} {\bibfnamefont {A.}~\bibnamefont
  {Schliesser}}, \bibinfo {author} {\bibfnamefont {R.}~\bibnamefont
  {Rivi\`{e}re}}, \bibinfo {author} {\bibfnamefont {G.}~\bibnamefont
  {Anetsberger}}, \bibinfo {author} {\bibfnamefont {O.}~\bibnamefont
  {Arcizet}},\ and\ \bibinfo {author} {\bibfnamefont {T.~J.}\ \bibnamefont
  {Kippenberg}},\ }\bibfield  {title} {\bibinfo {title} {Resolved-sideband
  cooling of a micromechanical oscillator},\ }\href
  {https://doi.org/10.1038/nphys939} {\bibfield  {journal} {\bibinfo  {journal}
  {Nat. Phys.}\ }\textbf {\bibinfo {volume} {4}},\ \bibinfo {pages} {415}
  (\bibinfo {year} {2008})}\BibitemShut {NoStop}%
\bibitem [{\citenamefont {Wilson-Rae}\ \emph {et~al.}(2008)\citenamefont
  {Wilson-Rae}, \citenamefont {Nooshi}, \citenamefont {Dobrindt}, \citenamefont
  {Kippenberg},\ and\ \citenamefont {Zwerger}}]{wilson2008cavity}%
  \BibitemOpen
  \bibfield  {author} {\bibinfo {author} {\bibfnamefont {I.}~\bibnamefont
  {Wilson-Rae}}, \bibinfo {author} {\bibfnamefont {N.}~\bibnamefont {Nooshi}},
  \bibinfo {author} {\bibfnamefont {J.}~\bibnamefont {Dobrindt}}, \bibinfo
  {author} {\bibfnamefont {T.~J.}\ \bibnamefont {Kippenberg}},\ and\ \bibinfo
  {author} {\bibfnamefont {W.}~\bibnamefont {Zwerger}},\ }\bibfield  {title}
  {\bibinfo {title} {Cavity-assisted backaction cooling of mechanical
  resonators},\ }\href {https://doi.org/10.1088/1367-2630/10/9/095007}
  {\bibfield  {journal} {\bibinfo  {journal} {New J. Phys.}\ }\textbf {\bibinfo
  {volume} {10}},\ \bibinfo {pages} {095007} (\bibinfo {year}
  {2008})}\BibitemShut {NoStop}%
\bibitem [{\citenamefont {JOUR}\ \emph {et~al.}(2010)\citenamefont {JOUR},
  \citenamefont {Rocheleau}, \citenamefont {Ndukum}, \citenamefont {Macklin},
  \citenamefont {Hertzberg}, \citenamefont {Clerk},\ and\ \citenamefont
  {Schwab}}]{Rocheleau2010}%
  \BibitemOpen
  \bibfield  {author} {\bibinfo {author} {\bibnamefont {JOUR}}, \bibinfo
  {author} {\bibfnamefont {T.}~\bibnamefont {Rocheleau}}, \bibinfo {author}
  {\bibfnamefont {T.}~\bibnamefont {Ndukum}}, \bibinfo {author} {\bibfnamefont
  {C.}~\bibnamefont {Macklin}}, \bibinfo {author} {\bibfnamefont {J.~B.}\
  \bibnamefont {Hertzberg}}, \bibinfo {author} {\bibfnamefont {A.~A.}\
  \bibnamefont {Clerk}},\ and\ \bibinfo {author} {\bibfnamefont {K.~C.}\
  \bibnamefont {Schwab}},\ }\bibfield  {title} {\bibinfo {title} {Preparation
  and detection of a mechanical resonator near the ground state of motion},\
  }\href {https://doi.org/10.1038/nature08681} {\bibfield  {journal} {\bibinfo
  {journal} {Nature}\ }\textbf {\bibinfo {volume} {463}},\ \bibinfo {pages}
  {72} (\bibinfo {year} {2010})}\BibitemShut {NoStop}%
\bibitem [{\citenamefont {O'Connell}\ \emph {et~al.}(2010)\citenamefont
  {O'Connell}, \citenamefont {Hofheinz}, \citenamefont {Ansmann}, \citenamefont
  {Bialczak}, \citenamefont {Lenander}, \citenamefont {Lucero}, \citenamefont
  {Neeley}, \citenamefont {Sank}, \citenamefont {Wang}, \citenamefont {Weides}
  \emph {et~al.}}]{o2010quantum}%
  \BibitemOpen
  \bibfield  {author} {\bibinfo {author} {\bibfnamefont {A.~D.}\ \bibnamefont
  {O'Connell}}, \bibinfo {author} {\bibfnamefont {M.}~\bibnamefont {Hofheinz}},
  \bibinfo {author} {\bibfnamefont {M.}~\bibnamefont {Ansmann}}, \bibinfo
  {author} {\bibfnamefont {R.~C.}\ \bibnamefont {Bialczak}}, \bibinfo {author}
  {\bibfnamefont {M.}~\bibnamefont {Lenander}}, \bibinfo {author}
  {\bibfnamefont {E.}~\bibnamefont {Lucero}}, \bibinfo {author} {\bibfnamefont
  {M.}~\bibnamefont {Neeley}}, \bibinfo {author} {\bibfnamefont
  {D.}~\bibnamefont {Sank}}, \bibinfo {author} {\bibfnamefont {H.}~\bibnamefont
  {Wang}}, \bibinfo {author} {\bibfnamefont {M.}~\bibnamefont {Weides}}, \emph
  {et~al.},\ }\bibfield  {title} {\bibinfo {title} {Quantum ground state and
  single-phonon control of a mechanical resonator},\ }\href
  {https://doi.org/10.1038/nature08967} {\bibfield  {journal} {\bibinfo
  {journal} {Nature}\ }\textbf {\bibinfo {volume} {464}},\ \bibinfo {pages}
  {697} (\bibinfo {year} {2010})}\BibitemShut {NoStop}%
\bibitem [{\citenamefont {Lai}\ \emph {et~al.}(2021)\citenamefont {Lai},
  \citenamefont {Huang}, \citenamefont {Hou}, \citenamefont {Nori},\ and\
  \citenamefont {Liao}}]{lai2021domino}%
  \BibitemOpen
  \bibfield  {author} {\bibinfo {author} {\bibfnamefont {D.-G.}\ \bibnamefont
  {Lai}}, \bibinfo {author} {\bibfnamefont {J.}~\bibnamefont {Huang}}, \bibinfo
  {author} {\bibfnamefont {B.-P.}\ \bibnamefont {Hou}}, \bibinfo {author}
  {\bibfnamefont {F.}~\bibnamefont {Nori}},\ and\ \bibinfo {author}
  {\bibfnamefont {J.-Q.}\ \bibnamefont {Liao}},\ }\bibfield  {title} {\bibinfo
  {title} {Domino cooling of a coupled mechanical-resonator chain via
  cold-damping feedback},\ }\href {https://doi.org/0} {\bibfield  {journal}
  {\bibinfo  {journal} {Phys. Rev. A}\ }\textbf {\bibinfo {volume} {103}},\
  \bibinfo {pages} {063509} (\bibinfo {year} {2021})}\BibitemShut {NoStop}%
\bibitem [{\citenamefont {Tao}\ \emph {et~al.}(2014)\citenamefont {Tao},
  \citenamefont {Boss}, \citenamefont {Moores},\ and\ \citenamefont
  {Degen}}]{Tao2014}%
  \BibitemOpen
  \bibfield  {author} {\bibinfo {author} {\bibfnamefont {Y.}~\bibnamefont
  {Tao}}, \bibinfo {author} {\bibfnamefont {J.~M.}\ \bibnamefont {Boss}},
  \bibinfo {author} {\bibfnamefont {B.~A.}\ \bibnamefont {Moores}},\ and\
  \bibinfo {author} {\bibfnamefont {C.~L.}\ \bibnamefont {Degen}},\ }\bibfield
  {title} {\bibinfo {title} {Single-crystal diamond nanomechanical resonators
  with quality factors exceeding one million},\ }\href
  {https://doi.org/10.1038/ncomms4638} {\bibfield  {journal} {\bibinfo
  {journal} {Nat. Commun.}\ }\textbf {\bibinfo {volume} {5}},\ \bibinfo {pages}
  {3638} (\bibinfo {year} {2014})}\BibitemShut {NoStop}%
\bibitem [{\citenamefont {Ovartchaiyapong}\ \emph {et~al.}(2012)\citenamefont
  {Ovartchaiyapong}, \citenamefont {Pascal}, \citenamefont {Myers},
  \citenamefont {Lauria},\ and\ \citenamefont
  {Bleszynski~Jayich}}]{doi:10.1063/1.4760274}%
  \BibitemOpen
  \bibfield  {author} {\bibinfo {author} {\bibfnamefont {P.}~\bibnamefont
  {Ovartchaiyapong}}, \bibinfo {author} {\bibfnamefont {L.~M.~A.}\ \bibnamefont
  {Pascal}}, \bibinfo {author} {\bibfnamefont {B.~A.}\ \bibnamefont {Myers}},
  \bibinfo {author} {\bibfnamefont {P.}~\bibnamefont {Lauria}},\ and\ \bibinfo
  {author} {\bibfnamefont {A.~C.}\ \bibnamefont {Bleszynski~Jayich}},\
  }\bibfield  {title} {\bibinfo {title} {High quality factor single-crystal
  diamond mechanical resonators},\ }\href {https://doi.org/10.1063/1.4760274}
  {\bibfield  {journal} {\bibinfo  {journal} {Appl. Phys. Lett.}\ }\textbf
  {\bibinfo {volume} {101}},\ \bibinfo {pages} {163505} (\bibinfo {year}
  {2012})}\BibitemShut {NoStop}%
\bibitem [{\citenamefont {Gaidarzhy}\ \emph {et~al.}(2007)\citenamefont
  {Gaidarzhy}, \citenamefont {Imboden}, \citenamefont {Mohanty}, \citenamefont
  {Rankin},\ and\ \citenamefont {Sheldon}}]{doi:10.1063/1.2804573}%
  \BibitemOpen
  \bibfield  {author} {\bibinfo {author} {\bibfnamefont {A.}~\bibnamefont
  {Gaidarzhy}}, \bibinfo {author} {\bibfnamefont {M.}~\bibnamefont {Imboden}},
  \bibinfo {author} {\bibfnamefont {P.}~\bibnamefont {Mohanty}}, \bibinfo
  {author} {\bibfnamefont {J.}~\bibnamefont {Rankin}},\ and\ \bibinfo {author}
  {\bibfnamefont {B.~W.}\ \bibnamefont {Sheldon}},\ }\bibfield  {title}
  {\bibinfo {title} {High quality factor gigahertz frequencies in
  nanomechanical diamond resonators},\ }\href
  {https://doi.org/10.1063/1.2804573} {\bibfield  {journal} {\bibinfo
  {journal} {Appl. Phys. Lett.}\ }\textbf {\bibinfo {volume} {91}},\ \bibinfo
  {pages} {203503} (\bibinfo {year} {2007})}\BibitemShut {NoStop}%
\bibitem [{\citenamefont {Aspelmeyer}\ \emph {et~al.}(2014)\citenamefont
  {Aspelmeyer}, \citenamefont {Kippenberg},\ and\ \citenamefont
  {Marquardt}}]{RevModPhys.86.1391}%
  \BibitemOpen
  \bibfield  {author} {\bibinfo {author} {\bibfnamefont {M.}~\bibnamefont
  {Aspelmeyer}}, \bibinfo {author} {\bibfnamefont {T.~J.}\ \bibnamefont
  {Kippenberg}},\ and\ \bibinfo {author} {\bibfnamefont {F.}~\bibnamefont
  {Marquardt}},\ }\bibfield  {title} {\bibinfo {title} {Cavity optomechanics},\
  }\href {https://doi.org/10.1103/RevModPhys.86.1391} {\bibfield  {journal}
  {\bibinfo  {journal} {Rev. Mod. Phys.}\ }\textbf {\bibinfo {volume} {86}},\
  \bibinfo {pages} {1391} (\bibinfo {year} {2014})}\BibitemShut {NoStop}%
\bibitem [{\citenamefont {Kippenberg}\ and\ \citenamefont
  {Vahala}(2008)}]{Kippenberg1172}%
  \BibitemOpen
  \bibfield  {author} {\bibinfo {author} {\bibfnamefont {T.~J.}\ \bibnamefont
  {Kippenberg}}\ and\ \bibinfo {author} {\bibfnamefont {K.~J.}\ \bibnamefont
  {Vahala}},\ }\bibfield  {title} {\bibinfo {title} {Cavity optomechanics:
  Back-action at the mesoscale},\ }\href
  {https://doi.org/10.1126/science.1156032} {\bibfield  {journal} {\bibinfo
  {journal} {Science}\ }\textbf {\bibinfo {volume} {321}},\ \bibinfo {pages}
  {1172} (\bibinfo {year} {2008})}\BibitemShut {NoStop}%
\bibitem [{\citenamefont {LaHaye}\ \emph {et~al.}(2009)\citenamefont {LaHaye},
  \citenamefont {Suh}, \citenamefont {Echternach}, \citenamefont {Schwab},\
  and\ \citenamefont {Roukes}}]{LaHaye2009}%
  \BibitemOpen
  \bibfield  {author} {\bibinfo {author} {\bibfnamefont {M.~D.}\ \bibnamefont
  {LaHaye}}, \bibinfo {author} {\bibfnamefont {J.}~\bibnamefont {Suh}},
  \bibinfo {author} {\bibfnamefont {P.~M.}\ \bibnamefont {Echternach}},
  \bibinfo {author} {\bibfnamefont {K.~C.}\ \bibnamefont {Schwab}},\ and\
  \bibinfo {author} {\bibfnamefont {M.~L.}\ \bibnamefont {Roukes}},\ }\bibfield
   {title} {\bibinfo {title} {Nanomechanical measurements of a superconducting
  qubit},\ }\href {https://doi.org/10.1038/nature08093} {\bibfield  {journal}
  {\bibinfo  {journal} {Nature}\ }\textbf {\bibinfo {volume} {459}},\ \bibinfo
  {pages} {960} (\bibinfo {year} {2009})}\BibitemShut {NoStop}%
\bibitem [{\citenamefont {J\"{o}ckel}\ \emph {et~al.}(2015)\citenamefont
  {J\"{o}ckel}, \citenamefont {Faber}, \citenamefont {Kampschulte},
  \citenamefont {Korppi}, \citenamefont {Rakher},\ and\ \citenamefont
  {Treutlein}}]{Jockel2015}%
  \BibitemOpen
  \bibfield  {author} {\bibinfo {author} {\bibfnamefont {A.}~\bibnamefont
  {J\"{o}ckel}}, \bibinfo {author} {\bibfnamefont {A.}~\bibnamefont {Faber}},
  \bibinfo {author} {\bibfnamefont {T.}~\bibnamefont {Kampschulte}}, \bibinfo
  {author} {\bibfnamefont {M.}~\bibnamefont {Korppi}}, \bibinfo {author}
  {\bibfnamefont {M.~T.}\ \bibnamefont {Rakher}},\ and\ \bibinfo {author}
  {\bibfnamefont {P.}~\bibnamefont {Treutlein}},\ }\bibfield  {title} {\bibinfo
  {title} {Sympathetic cooling of a membrane oscillator in a hybrid
  mechanical–atomic system},\ }\href {https://doi.org/10.1038/nnano.2014.278}
  {\bibfield  {journal} {\bibinfo  {journal} {Nat. Nanotechnol.}\ }\textbf
  {\bibinfo {volume} {10}},\ \bibinfo {pages} {55} (\bibinfo {year}
  {2015})}\BibitemShut {NoStop}%
\bibitem [{\citenamefont {Montinaro}\ \emph {et~al.}(2014)\citenamefont
  {Montinaro}, \citenamefont {W\"{u}st}, \citenamefont {Munsch}, \citenamefont
  {Fontana}, \citenamefont {Russo-Averchi}, \citenamefont {Heiss},
  \citenamefont {Fontcuberta~i Morral}, \citenamefont {Warburton},\ and\
  \citenamefont {Poggio}}]{doi:10.1021/nl501413t}%
  \BibitemOpen
  \bibfield  {author} {\bibinfo {author} {\bibfnamefont {M.}~\bibnamefont
  {Montinaro}}, \bibinfo {author} {\bibfnamefont {G.}~\bibnamefont {W\"{u}st}},
  \bibinfo {author} {\bibfnamefont {M.}~\bibnamefont {Munsch}}, \bibinfo
  {author} {\bibfnamefont {Y.}~\bibnamefont {Fontana}}, \bibinfo {author}
  {\bibfnamefont {E.}~\bibnamefont {Russo-Averchi}}, \bibinfo {author}
  {\bibfnamefont {M.}~\bibnamefont {Heiss}}, \bibinfo {author} {\bibfnamefont
  {A.}~\bibnamefont {Fontcuberta~i Morral}}, \bibinfo {author} {\bibfnamefont
  {R.~J.}\ \bibnamefont {Warburton}},\ and\ \bibinfo {author} {\bibfnamefont
  {M.}~\bibnamefont {Poggio}},\ }\bibfield  {title} {\bibinfo {title} {Quantum
  dot opto-mechanics in a fully self-assembled nanowire},\ }\href
  {https://doi.org/10.1021/nl501413t} {\bibfield  {journal} {\bibinfo
  {journal} {Nano. Lett.}\ }\textbf {\bibinfo {volume} {14}},\ \bibinfo {pages}
  {4454} (\bibinfo {year} {2014})}\BibitemShut {NoStop}%
\bibitem [{\citenamefont {Yeo}\ \emph {et~al.}(2014)\citenamefont {Yeo},
  \citenamefont {de~Assis}, \citenamefont {Gloppe}, \citenamefont
  {Dupont-Ferrier}, \citenamefont {Verlot}, \citenamefont {Malik},
  \citenamefont {Dupuy}, \citenamefont {Claudon}, \citenamefont {G\'{e}rard},
  \citenamefont {Auff\`{e}ves}, \citenamefont {Nogues}, \citenamefont
  {Seidelin}, \citenamefont {Poizat}, \citenamefont {Arcizet},\ and\
  \citenamefont {Richard}}]{Yeo2014}%
  \BibitemOpen
  \bibfield  {author} {\bibinfo {author} {\bibfnamefont {I.}~\bibnamefont
  {Yeo}}, \bibinfo {author} {\bibfnamefont {P.-L.}\ \bibnamefont {de~Assis}},
  \bibinfo {author} {\bibfnamefont {A.}~\bibnamefont {Gloppe}}, \bibinfo
  {author} {\bibfnamefont {E.}~\bibnamefont {Dupont-Ferrier}}, \bibinfo
  {author} {\bibfnamefont {P.}~\bibnamefont {Verlot}}, \bibinfo {author}
  {\bibfnamefont {N.~S.}\ \bibnamefont {Malik}}, \bibinfo {author}
  {\bibfnamefont {E.}~\bibnamefont {Dupuy}}, \bibinfo {author} {\bibfnamefont
  {J.}~\bibnamefont {Claudon}}, \bibinfo {author} {\bibfnamefont {J.-M.}\
  \bibnamefont {G\'{e}rard}}, \bibinfo {author} {\bibfnamefont
  {A.}~\bibnamefont {Auff\`{e}ves}}, \bibinfo {author} {\bibfnamefont
  {G.}~\bibnamefont {Nogues}}, \bibinfo {author} {\bibfnamefont
  {S.}~\bibnamefont {Seidelin}}, \bibinfo {author} {\bibfnamefont {J.-P.}\
  \bibnamefont {Poizat}}, \bibinfo {author} {\bibfnamefont {O.}~\bibnamefont
  {Arcizet}},\ and\ \bibinfo {author} {\bibfnamefont {M.}~\bibnamefont
  {Richard}},\ }\bibfield  {title} {\bibinfo {title} {Strain-mediated coupling
  in a quantum dot–mechanical oscillator hybrid system},\ }\href
  {https://doi.org/10.1038/nnano.2013.274} {\bibfield  {journal} {\bibinfo
  {journal} {Nat. Nanotechnol.}\ }\textbf {\bibinfo {volume} {9}},\ \bibinfo
  {pages} {106} (\bibinfo {year} {2014})}\BibitemShut {NoStop}%
\bibitem [{\citenamefont {Kolkowitz}\ \emph {et~al.}(2012)\citenamefont
  {Kolkowitz}, \citenamefont {Bleszynski~Jayich}, \citenamefont
  {Unterreithmeier}, \citenamefont {Bennett}, \citenamefont {Rabl},
  \citenamefont {Harris},\ and\ \citenamefont {Lukin}}]{Kolkowitz1603}%
  \BibitemOpen
  \bibfield  {author} {\bibinfo {author} {\bibfnamefont {S.}~\bibnamefont
  {Kolkowitz}}, \bibinfo {author} {\bibfnamefont {A.~C.}\ \bibnamefont
  {Bleszynski~Jayich}}, \bibinfo {author} {\bibfnamefont {Q.~P.}\ \bibnamefont
  {Unterreithmeier}}, \bibinfo {author} {\bibfnamefont {S.~D.}\ \bibnamefont
  {Bennett}}, \bibinfo {author} {\bibfnamefont {P.}~\bibnamefont {Rabl}},
  \bibinfo {author} {\bibfnamefont {J.~G.~E.}\ \bibnamefont {Harris}},\ and\
  \bibinfo {author} {\bibfnamefont {M.~D.}\ \bibnamefont {Lukin}},\ }\bibfield
  {title} {\bibinfo {title} {Coherent sensing of a mechanical resonator with a
  single-spin qubit},\ }\href {https://doi.org/10.1126/science.1216821}
  {\bibfield  {journal} {\bibinfo  {journal} {Science}\ }\textbf {\bibinfo
  {volume} {335}},\ \bibinfo {pages} {1603} (\bibinfo {year}
  {2012})}\BibitemShut {NoStop}%
\bibitem [{\citenamefont {Kuzyk}\ and\ \citenamefont
  {Wang}(2018)}]{PhysRevX.8.041027}%
  \BibitemOpen
  \bibfield  {author} {\bibinfo {author} {\bibfnamefont {M.~C.}\ \bibnamefont
  {Kuzyk}}\ and\ \bibinfo {author} {\bibfnamefont {H.}~\bibnamefont {Wang}},\
  }\bibfield  {title} {\bibinfo {title} {Scaling phononic quantum networks of
  solid-state spins with closed mechanical subsystems},\ }\href
  {https://doi.org/10.1103/PhysRevX.8.041027} {\bibfield  {journal} {\bibinfo
  {journal} {Phys. Rev. X}\ }\textbf {\bibinfo {volume} {8}},\ \bibinfo {pages}
  {041027} (\bibinfo {year} {2018})}\BibitemShut {NoStop}%
\bibitem [{\citenamefont {Lemonde}\ \emph {et~al.}(2018)\citenamefont
  {Lemonde}, \citenamefont {Meesala}, \citenamefont {Sipahigil}, \citenamefont
  {Schuetz}, \citenamefont {Lukin}, \citenamefont {Loncar},\ and\ \citenamefont
  {Rabl}}]{PhysRevLett.120.213603}%
  \BibitemOpen
  \bibfield  {author} {\bibinfo {author} {\bibfnamefont {M.-A.}\ \bibnamefont
  {Lemonde}}, \bibinfo {author} {\bibfnamefont {S.}~\bibnamefont {Meesala}},
  \bibinfo {author} {\bibfnamefont {A.}~\bibnamefont {Sipahigil}}, \bibinfo
  {author} {\bibfnamefont {M.~J.~A.}\ \bibnamefont {Schuetz}}, \bibinfo
  {author} {\bibfnamefont {M.~D.}\ \bibnamefont {Lukin}}, \bibinfo {author}
  {\bibfnamefont {M.}~\bibnamefont {Loncar}},\ and\ \bibinfo {author}
  {\bibfnamefont {P.}~\bibnamefont {Rabl}},\ }\bibfield  {title} {\bibinfo
  {title} {Phonon networks with silicon-vacancy centers in diamond
  waveguides},\ }\href {https://doi.org/10.1103/PhysRevLett.120.213603}
  {\bibfield  {journal} {\bibinfo  {journal} {Phys. Rev. Lett.}\ }\textbf
  {\bibinfo {volume} {120}},\ \bibinfo {pages} {213603} (\bibinfo {year}
  {2018})}\BibitemShut {NoStop}%
\bibitem [{\citenamefont {Peano}\ \emph {et~al.}(2015)\citenamefont {Peano},
  \citenamefont {Brendel}, \citenamefont {Schmidt},\ and\ \citenamefont
  {Marquardt}}]{PhysRevX.5.031011}%
  \BibitemOpen
  \bibfield  {author} {\bibinfo {author} {\bibfnamefont {V.}~\bibnamefont
  {Peano}}, \bibinfo {author} {\bibfnamefont {C.}~\bibnamefont {Brendel}},
  \bibinfo {author} {\bibfnamefont {M.}~\bibnamefont {Schmidt}},\ and\ \bibinfo
  {author} {\bibfnamefont {F.}~\bibnamefont {Marquardt}},\ }\bibfield  {title}
  {\bibinfo {title} {Topological phases of sound and light},\ }\href
  {https://doi.org/10.1103/PhysRevX.5.031011} {\bibfield  {journal} {\bibinfo
  {journal} {Phys. Rev. X}\ }\textbf {\bibinfo {volume} {5}},\ \bibinfo {pages}
  {031011} (\bibinfo {year} {2015})}\BibitemShut {NoStop}%
\bibitem [{\citenamefont {Lemonde}\ \emph {et~al.}(2019)\citenamefont
  {Lemonde}, \citenamefont {Peano}, \citenamefont {Rabl},\ and\ \citenamefont
  {Angelakis}}]{Lemonde_2019}%
  \BibitemOpen
  \bibfield  {author} {\bibinfo {author} {\bibfnamefont {M.-A.}\ \bibnamefont
  {Lemonde}}, \bibinfo {author} {\bibfnamefont {V.}~\bibnamefont {Peano}},
  \bibinfo {author} {\bibfnamefont {P.}~\bibnamefont {Rabl}},\ and\ \bibinfo
  {author} {\bibfnamefont {D.~G.}\ \bibnamefont {Angelakis}},\ }\bibfield
  {title} {\bibinfo {title} {Quantum state transfer via acoustic edge states in
  a 2d optomechanical array},\ }\href
  {https://doi.org/10.1088/1367-2630/ab51f5} {\bibfield  {journal} {\bibinfo
  {journal} {New J. Phys.}\ }\textbf {\bibinfo {volume} {21}},\ \bibinfo
  {pages} {113030} (\bibinfo {year} {2019})}\BibitemShut {NoStop}%
\bibitem [{\citenamefont {Liu}\ \emph {et~al.}(2010)\citenamefont {Liu},
  \citenamefont {Miranowicz}, \citenamefont {Gao}, \citenamefont {Bajer},
  \citenamefont {Sun},\ and\ \citenamefont {Nori}}]{PhysRevA.82.032101}%
  \BibitemOpen
  \bibfield  {author} {\bibinfo {author} {\bibfnamefont {Y.-x.}\ \bibnamefont
  {Liu}}, \bibinfo {author} {\bibfnamefont {A.}~\bibnamefont {Miranowicz}},
  \bibinfo {author} {\bibfnamefont {Y.~B.}\ \bibnamefont {Gao}}, \bibinfo
  {author} {\bibfnamefont {J.~c.~v.}\ \bibnamefont {Bajer}}, \bibinfo {author}
  {\bibfnamefont {C.~P.}\ \bibnamefont {Sun}},\ and\ \bibinfo {author}
  {\bibfnamefont {F.}~\bibnamefont {Nori}},\ }\bibfield  {title} {\bibinfo
  {title} {Qubit-induced phonon blockade as a signature of quantum behavior in
  nanomechanical resonators},\ }\href
  {https://doi.org/10.1103/PhysRevA.82.032101} {\bibfield  {journal} {\bibinfo
  {journal} {Phys. Rev. A}\ }\textbf {\bibinfo {volume} {82}},\ \bibinfo
  {pages} {032101} (\bibinfo {year} {2010})}\BibitemShut {NoStop}%
\bibitem [{\citenamefont {Huang}\ \emph {et~al.}(2018)\citenamefont {Huang},
  \citenamefont {Miranowicz}, \citenamefont {Liao}, \citenamefont {Nori},\ and\
  \citenamefont {Jing}}]{PhysRevLett.121.153601}%
  \BibitemOpen
  \bibfield  {author} {\bibinfo {author} {\bibfnamefont {R.}~\bibnamefont
  {Huang}}, \bibinfo {author} {\bibfnamefont {A.}~\bibnamefont {Miranowicz}},
  \bibinfo {author} {\bibfnamefont {J.-Q.}\ \bibnamefont {Liao}}, \bibinfo
  {author} {\bibfnamefont {F.}~\bibnamefont {Nori}},\ and\ \bibinfo {author}
  {\bibfnamefont {H.}~\bibnamefont {Jing}},\ }\bibfield  {title} {\bibinfo
  {title} {Nonreciprocal photon blockade},\ }\href
  {https://doi.org/10.1103/PhysRevLett.121.153601} {\bibfield  {journal}
  {\bibinfo  {journal} {Phys. Rev. Lett.}\ }\textbf {\bibinfo {volume} {121}},\
  \bibinfo {pages} {153601} (\bibinfo {year} {2018})}\BibitemShut {NoStop}%
\bibitem [{\citenamefont {Zheng}\ \emph {et~al.}(2019)\citenamefont {Zheng},
  \citenamefont {Yin}, \citenamefont {Bin}, \citenamefont {L\"u},\ and\
  \citenamefont {Wu}}]{PhysRevA.99.013804}%
  \BibitemOpen
  \bibfield  {author} {\bibinfo {author} {\bibfnamefont {L.-L.}\ \bibnamefont
  {Zheng}}, \bibinfo {author} {\bibfnamefont {T.-S.}\ \bibnamefont {Yin}},
  \bibinfo {author} {\bibfnamefont {Q.}~\bibnamefont {Bin}}, \bibinfo {author}
  {\bibfnamefont {X.-Y.}\ \bibnamefont {L\"u}},\ and\ \bibinfo {author}
  {\bibfnamefont {Y.}~\bibnamefont {Wu}},\ }\bibfield  {title} {\bibinfo
  {title} {Single-photon-induced phonon blockade in a hybrid
  spin-optomechanical system},\ }\href
  {https://doi.org/10.1103/PhysRevA.99.013804} {\bibfield  {journal} {\bibinfo
  {journal} {Phys. Rev. A}\ }\textbf {\bibinfo {volume} {99}},\ \bibinfo
  {pages} {013804} (\bibinfo {year} {2019})}\BibitemShut {NoStop}%
\bibitem [{\citenamefont {Wang}\ \emph {et~al.}(2016)\citenamefont {Wang},
  \citenamefont {Miranowicz}, \citenamefont {Li},\ and\ \citenamefont
  {Nori}}]{PhysRevA.93.063861}%
  \BibitemOpen
  \bibfield  {author} {\bibinfo {author} {\bibfnamefont {X.}~\bibnamefont
  {Wang}}, \bibinfo {author} {\bibfnamefont {A.}~\bibnamefont {Miranowicz}},
  \bibinfo {author} {\bibfnamefont {H.-R.}\ \bibnamefont {Li}},\ and\ \bibinfo
  {author} {\bibfnamefont {F.}~\bibnamefont {Nori}},\ }\bibfield  {title}
  {\bibinfo {title} {Method for observing robust and tunable phonon blockade in
  a nanomechanical resonator coupled to a charge qubit},\ }\href
  {https://doi.org/10.1103/PhysRevA.93.063861} {\bibfield  {journal} {\bibinfo
  {journal} {Phys. Rev. A}\ }\textbf {\bibinfo {volume} {93}},\ \bibinfo
  {pages} {063861} (\bibinfo {year} {2016})}\BibitemShut {NoStop}%
\bibitem [{\citenamefont {Cai}\ \emph {et~al.}(2018)\citenamefont {Cai},
  \citenamefont {Pan}, \citenamefont {Wang}, \citenamefont {Ruan},
  \citenamefont {Yin},\ and\ \citenamefont {Long}}]{Cai:18}%
  \BibitemOpen
  \bibfield  {author} {\bibinfo {author} {\bibfnamefont {K.}~\bibnamefont
  {Cai}}, \bibinfo {author} {\bibfnamefont {Z.-W.}\ \bibnamefont {Pan}},
  \bibinfo {author} {\bibfnamefont {R.-X.}\ \bibnamefont {Wang}}, \bibinfo
  {author} {\bibfnamefont {D.}~\bibnamefont {Ruan}}, \bibinfo {author}
  {\bibfnamefont {Z.-Q.}\ \bibnamefont {Yin}},\ and\ \bibinfo {author}
  {\bibfnamefont {G.-L.}\ \bibnamefont {Long}},\ }\bibfield  {title} {\bibinfo
  {title} {Single phonon source based on a giant polariton nonlinear effect},\
  }\href {https://doi.org/10.1364/OL.43.001163} {\bibfield  {journal} {\bibinfo
   {journal} {Opt. Lett.}\ }\textbf {\bibinfo {volume} {43}},\ \bibinfo {pages}
  {1163} (\bibinfo {year} {2018})}\BibitemShut {NoStop}%
\bibitem [{\citenamefont {Doherty}\ \emph {et~al.}(2014)\citenamefont
  {Doherty}, \citenamefont {Struzhkin}, \citenamefont {Simpson}, \citenamefont
  {McGuinness}, \citenamefont {Meng}, \citenamefont {Stacey}, \citenamefont
  {Karle}, \citenamefont {Hemley}, \citenamefont {Manson}, \citenamefont
  {Hollenberg},\ and\ \citenamefont {Prawer}}]{PhysRevLett.112.047601}%
  \BibitemOpen
  \bibfield  {author} {\bibinfo {author} {\bibfnamefont {M.~W.}\ \bibnamefont
  {Doherty}}, \bibinfo {author} {\bibfnamefont {V.~V.}\ \bibnamefont
  {Struzhkin}}, \bibinfo {author} {\bibfnamefont {D.~A.}\ \bibnamefont
  {Simpson}}, \bibinfo {author} {\bibfnamefont {L.~P.}\ \bibnamefont
  {McGuinness}}, \bibinfo {author} {\bibfnamefont {Y.}~\bibnamefont {Meng}},
  \bibinfo {author} {\bibfnamefont {A.}~\bibnamefont {Stacey}}, \bibinfo
  {author} {\bibfnamefont {T.~J.}\ \bibnamefont {Karle}}, \bibinfo {author}
  {\bibfnamefont {R.~J.}\ \bibnamefont {Hemley}}, \bibinfo {author}
  {\bibfnamefont {N.~B.}\ \bibnamefont {Manson}}, \bibinfo {author}
  {\bibfnamefont {L.~C.~L.}\ \bibnamefont {Hollenberg}},\ and\ \bibinfo
  {author} {\bibfnamefont {S.}~\bibnamefont {Prawer}},\ }\bibfield  {title}
  {\bibinfo {title} {Electronic properties and metrology applications of the
  diamond ${\mathrm{nv}}^{\ensuremath{-}}$ center under pressure},\ }\href
  {https://doi.org/10.1103/PhysRevLett.112.047601} {\bibfield  {journal}
  {\bibinfo  {journal} {Phys. Rev. Lett.}\ }\textbf {\bibinfo {volume} {112}},\
  \bibinfo {pages} {047601} (\bibinfo {year} {2014})}\BibitemShut {NoStop}%
\bibitem [{\citenamefont {Li}\ \emph {et~al.}(2016)\citenamefont {Li},
  \citenamefont {Xiang}, \citenamefont {Rabl},\ and\ \citenamefont
  {Nori}}]{PhysRevLett.117.015502}%
  \BibitemOpen
  \bibfield  {author} {\bibinfo {author} {\bibfnamefont {P.-B.}\ \bibnamefont
  {Li}}, \bibinfo {author} {\bibfnamefont {Z.-L.}\ \bibnamefont {Xiang}},
  \bibinfo {author} {\bibfnamefont {P.}~\bibnamefont {Rabl}},\ and\ \bibinfo
  {author} {\bibfnamefont {F.}~\bibnamefont {Nori}},\ }\bibfield  {title}
  {\bibinfo {title} {Hybrid quantum device with nitrogen-vacancy centers in
  diamond coupled to carbon nanotubes},\ }\href
  {https://doi.org/10.1103/PhysRevLett.117.015502} {\bibfield  {journal}
  {\bibinfo  {journal} {Phys. Rev. Lett.}\ }\textbf {\bibinfo {volume} {117}},\
  \bibinfo {pages} {015502} (\bibinfo {year} {2016})}\BibitemShut {NoStop}%
\bibitem [{\citenamefont {Golter}\ \emph
  {et~al.}(2016{\natexlab{a}})\citenamefont {Golter}, \citenamefont {Oo},
  \citenamefont {Amezcua}, \citenamefont {Lekavicius}, \citenamefont
  {Stewart},\ and\ \citenamefont {Wang}}]{PhysRevX.6.041060}%
  \BibitemOpen
  \bibfield  {author} {\bibinfo {author} {\bibfnamefont {D.~A.}\ \bibnamefont
  {Golter}}, \bibinfo {author} {\bibfnamefont {T.}~\bibnamefont {Oo}}, \bibinfo
  {author} {\bibfnamefont {M.}~\bibnamefont {Amezcua}}, \bibinfo {author}
  {\bibfnamefont {I.}~\bibnamefont {Lekavicius}}, \bibinfo {author}
  {\bibfnamefont {K.~A.}\ \bibnamefont {Stewart}},\ and\ \bibinfo {author}
  {\bibfnamefont {H.}~\bibnamefont {Wang}},\ }\bibfield  {title} {\bibinfo
  {title} {Coupling a surface acoustic wave to an electron spin in diamond via
  a dark state},\ }\href {https://doi.org/10.1103/PhysRevX.6.041060} {\bibfield
   {journal} {\bibinfo  {journal} {Phys. Rev. X}\ }\textbf {\bibinfo {volume}
  {6}},\ \bibinfo {pages} {041060} (\bibinfo {year}
  {2016}{\natexlab{a}})}\BibitemShut {NoStop}%
\bibitem [{\citenamefont {Golter}\ \emph
  {et~al.}(2016{\natexlab{b}})\citenamefont {Golter}, \citenamefont {Oo},
  \citenamefont {Amezcua}, \citenamefont {Stewart},\ and\ \citenamefont
  {Wang}}]{PhysRevLett.116.143602}%
  \BibitemOpen
  \bibfield  {author} {\bibinfo {author} {\bibfnamefont {D.~A.}\ \bibnamefont
  {Golter}}, \bibinfo {author} {\bibfnamefont {T.}~\bibnamefont {Oo}}, \bibinfo
  {author} {\bibfnamefont {M.}~\bibnamefont {Amezcua}}, \bibinfo {author}
  {\bibfnamefont {K.~A.}\ \bibnamefont {Stewart}},\ and\ \bibinfo {author}
  {\bibfnamefont {H.}~\bibnamefont {Wang}},\ }\bibfield  {title} {\bibinfo
  {title} {Optomechanical quantum control of a nitrogen-vacancy center in
  diamond},\ }\href {https://doi.org/10.1103/PhysRevLett.116.143602} {\bibfield
   {journal} {\bibinfo  {journal} {Phys. Rev. Lett.}\ }\textbf {\bibinfo
  {volume} {116}},\ \bibinfo {pages} {143602} (\bibinfo {year}
  {2016}{\natexlab{b}})}\BibitemShut {NoStop}%
\bibitem [{\citenamefont {Zhou}\ \emph {et~al.}(2017)\citenamefont {Zhou},
  \citenamefont {Ma}, \citenamefont {Li}, \citenamefont {Li}, \citenamefont
  {Li},\ and\ \citenamefont {Li}}]{PhysRevA.96.062333}%
  \BibitemOpen
  \bibfield  {author} {\bibinfo {author} {\bibfnamefont {Y.}~\bibnamefont
  {Zhou}}, \bibinfo {author} {\bibfnamefont {S.-L.}\ \bibnamefont {Ma}},
  \bibinfo {author} {\bibfnamefont {B.}~\bibnamefont {Li}}, \bibinfo {author}
  {\bibfnamefont {X.-X.}\ \bibnamefont {Li}}, \bibinfo {author} {\bibfnamefont
  {F.-L.}\ \bibnamefont {Li}},\ and\ \bibinfo {author} {\bibfnamefont {P.-B.}\
  \bibnamefont {Li}},\ }\bibfield  {title} {\bibinfo {title} {Simulating the
  {Lipkin-Meshkov-Glick} model in a hybrid quantum system},\ }\href
  {https://doi.org/10.1103/PhysRevA.96.062333} {\bibfield  {journal} {\bibinfo
  {journal} {Phys. Rev. A}\ }\textbf {\bibinfo {volume} {96}},\ \bibinfo
  {pages} {062333} (\bibinfo {year} {2017})}\BibitemShut {NoStop}%
\bibitem [{\citenamefont {Li}\ and\ \citenamefont
  {Nori}(2018)}]{PhysRevApplied.10.024011}%
  \BibitemOpen
  \bibfield  {author} {\bibinfo {author} {\bibfnamefont {P.-B.}\ \bibnamefont
  {Li}}\ and\ \bibinfo {author} {\bibfnamefont {F.}~\bibnamefont {Nori}},\
  }\bibfield  {title} {\bibinfo {title} {Hybrid quantum system with
  nitrogen-vacancy centers in diamond coupled to surface-phonon polaritons in
  piezomagnetic superlattices},\ }\href
  {https://doi.org/10.1103/PhysRevApplied.10.024011} {\bibfield  {journal}
  {\bibinfo  {journal} {Phys. Rev. Appl.}\ }\textbf {\bibinfo {volume} {10}},\
  \bibinfo {pages} {024011} (\bibinfo {year} {2018})}\BibitemShut {NoStop}%
\bibitem [{\citenamefont {Li}\ \emph {et~al.}(2015)\citenamefont {Li},
  \citenamefont {Liu}, \citenamefont {Gao}, \citenamefont {Xiang},
  \citenamefont {Rabl}, \citenamefont {Xiao},\ and\ \citenamefont
  {Li}}]{PhysRevApplied.4.044003}%
  \BibitemOpen
  \bibfield  {author} {\bibinfo {author} {\bibfnamefont {P.-B.}\ \bibnamefont
  {Li}}, \bibinfo {author} {\bibfnamefont {Y.-C.}\ \bibnamefont {Liu}},
  \bibinfo {author} {\bibfnamefont {S.-Y.}\ \bibnamefont {Gao}}, \bibinfo
  {author} {\bibfnamefont {Z.-L.}\ \bibnamefont {Xiang}}, \bibinfo {author}
  {\bibfnamefont {P.}~\bibnamefont {Rabl}}, \bibinfo {author} {\bibfnamefont
  {Y.-F.}\ \bibnamefont {Xiao}},\ and\ \bibinfo {author} {\bibfnamefont
  {F.-L.}\ \bibnamefont {Li}},\ }\bibfield  {title} {\bibinfo {title} {Hybrid
  quantum device based on {NV} centers in diamond nanomechanical resonators
  plus superconducting waveguide cavities},\ }\href
  {https://doi.org/10.1103/PhysRevApplied.4.044003} {\bibfield  {journal}
  {\bibinfo  {journal} {Phys. Rev. Appl.}\ }\textbf {\bibinfo {volume} {4}},\
  \bibinfo {pages} {044003} (\bibinfo {year} {2015})}\BibitemShut {NoStop}%
\bibitem [{\citenamefont {Ai}\ \emph {et~al.}(2021)\citenamefont {Ai},
  \citenamefont {Li}, \citenamefont {Qin}, \citenamefont {Zhao}, \citenamefont
  {Sun}, \citenamefont {Nori} \emph {et~al.}}]{ai2021nv}%
  \BibitemOpen
  \bibfield  {author} {\bibinfo {author} {\bibfnamefont {Q.}~\bibnamefont
  {Ai}}, \bibinfo {author} {\bibfnamefont {P.-B.}\ \bibnamefont {Li}}, \bibinfo
  {author} {\bibfnamefont {W.}~\bibnamefont {Qin}}, \bibinfo {author}
  {\bibfnamefont {J.-X.}\ \bibnamefont {Zhao}}, \bibinfo {author}
  {\bibfnamefont {C.}~\bibnamefont {Sun}}, \bibinfo {author} {\bibfnamefont
  {F.}~\bibnamefont {Nori}}, \emph {et~al.},\ }\bibfield  {title} {\bibinfo
  {title} {The \text{NV} metamaterial: Tunable quantum hyperbolic metamaterial
  using nitrogen vacancy centers in diamond},\ }\href {https://doi.org/0}
  {\bibfield  {journal} {\bibinfo  {journal} {Phys. Rev. B}\ }\textbf {\bibinfo
  {volume} {104}},\ \bibinfo {pages} {014109} (\bibinfo {year}
  {2021})}\BibitemShut {NoStop}%
\bibitem [{\citenamefont {Xiong}\ \emph {et~al.}(2021)\citenamefont {Xiong},
  \citenamefont {Chen}, \citenamefont {Fang}, \citenamefont {Wang},
  \citenamefont {Ye},\ and\ \citenamefont {You}}]{xiong2021strong}%
  \BibitemOpen
  \bibfield  {author} {\bibinfo {author} {\bibfnamefont {W.}~\bibnamefont
  {Xiong}}, \bibinfo {author} {\bibfnamefont {J.}~\bibnamefont {Chen}},
  \bibinfo {author} {\bibfnamefont {B.}~\bibnamefont {Fang}}, \bibinfo {author}
  {\bibfnamefont {M.}~\bibnamefont {Wang}}, \bibinfo {author} {\bibfnamefont
  {L.}~\bibnamefont {Ye}},\ and\ \bibinfo {author} {\bibfnamefont
  {J.}~\bibnamefont {You}},\ }\bibfield  {title} {\bibinfo {title} {Strong
  tunable spin-spin interaction in a weakly coupled nitrogen vacancy
  spin-cavity electromechanical system},\ }\href {https://doi.org/0} {\bibfield
   {journal} {\bibinfo  {journal} {Phys. Rev. B}\ }\textbf {\bibinfo {volume}
  {103}},\ \bibinfo {pages} {174106} (\bibinfo {year} {2021})}\BibitemShut
  {NoStop}%
\bibitem [{\citenamefont {Hei}\ \emph {et~al.}(2021)\citenamefont {Hei},
  \citenamefont {Dong}, \citenamefont {Chen}, \citenamefont {Shen},
  \citenamefont {Qiao},\ and\ \citenamefont {Li}}]{hei2021enhancing}%
  \BibitemOpen
  \bibfield  {author} {\bibinfo {author} {\bibfnamefont {X.-L.}\ \bibnamefont
  {Hei}}, \bibinfo {author} {\bibfnamefont {X.-L.}\ \bibnamefont {Dong}},
  \bibinfo {author} {\bibfnamefont {J.-Q.}\ \bibnamefont {Chen}}, \bibinfo
  {author} {\bibfnamefont {C.-P.}\ \bibnamefont {Shen}}, \bibinfo {author}
  {\bibfnamefont {Y.-F.}\ \bibnamefont {Qiao}},\ and\ \bibinfo {author}
  {\bibfnamefont {P.-B.}\ \bibnamefont {Li}},\ }\bibfield  {title} {\bibinfo
  {title} {Enhancing spin-photon coupling with a micromagnet},\ }\href
  {https://doi.org/0} {\bibfield  {journal} {\bibinfo  {journal} {Phys. Rev.
  A}\ }\textbf {\bibinfo {volume} {103}},\ \bibinfo {pages} {043706} (\bibinfo
  {year} {2021})}\BibitemShut {NoStop}%
\bibitem [{\citenamefont {Zhou}\ \emph {et~al.}(2021)\citenamefont {Zhou},
  \citenamefont {L{\"u}},\ and\ \citenamefont {Zeng}}]{zhou2021chiral}%
  \BibitemOpen
  \bibfield  {author} {\bibinfo {author} {\bibfnamefont {Y.}~\bibnamefont
  {Zhou}}, \bibinfo {author} {\bibfnamefont {D.-Y.}\ \bibnamefont {L{\"u}}},\
  and\ \bibinfo {author} {\bibfnamefont {W.-Y.}\ \bibnamefont {Zeng}},\
  }\bibfield  {title} {\bibinfo {title} {Chiral single-photon switch-assisted
  quantum logic gate with a nitrogen-vacancy center in a hybrid system},\
  }\href {https://doi.org/0} {\bibfield  {journal} {\bibinfo  {journal}
  {Photonics Research}\ }\textbf {\bibinfo {volume} {9}},\ \bibinfo {pages}
  {405} (\bibinfo {year} {2021})}\BibitemShut {NoStop}%
\bibitem [{\citenamefont {Huillery}\ \emph {et~al.}(2020)\citenamefont
  {Huillery}, \citenamefont {Delord}, \citenamefont {Nicolas}, \citenamefont
  {Van Den~Bossche}, \citenamefont {Perdriat},\ and\ \citenamefont
  {Hetet}}]{huillery2020spin}%
  \BibitemOpen
  \bibfield  {author} {\bibinfo {author} {\bibfnamefont {P.}~\bibnamefont
  {Huillery}}, \bibinfo {author} {\bibfnamefont {T.}~\bibnamefont {Delord}},
  \bibinfo {author} {\bibfnamefont {L.}~\bibnamefont {Nicolas}}, \bibinfo
  {author} {\bibfnamefont {M.}~\bibnamefont {Van Den~Bossche}}, \bibinfo
  {author} {\bibfnamefont {M.}~\bibnamefont {Perdriat}},\ and\ \bibinfo
  {author} {\bibfnamefont {G.}~\bibnamefont {Hetet}},\ }\bibfield  {title}
  {\bibinfo {title} {Spin mechanics with levitating ferromagnetic particles},\
  }\href {https://doi.org/0} {\bibfield  {journal} {\bibinfo  {journal} {Phys.
  Rev. B}\ }\textbf {\bibinfo {volume} {101}},\ \bibinfo {pages} {134415}
  (\bibinfo {year} {2020})}\BibitemShut {NoStop}%
\bibitem [{\citenamefont {Bar-Gill}\ \emph {et~al.}(2013)\citenamefont
  {Bar-Gill}, \citenamefont {Pham}, \citenamefont {Jarmola}, \citenamefont
  {Budker},\ and\ \citenamefont {Walsworth}}]{Bar-Gill2013}%
  \BibitemOpen
  \bibfield  {author} {\bibinfo {author} {\bibfnamefont {N.}~\bibnamefont
  {Bar-Gill}}, \bibinfo {author} {\bibfnamefont {L.}~\bibnamefont {Pham}},
  \bibinfo {author} {\bibfnamefont {A.}~\bibnamefont {Jarmola}}, \bibinfo
  {author} {\bibfnamefont {D.}~\bibnamefont {Budker}},\ and\ \bibinfo {author}
  {\bibfnamefont {R.}~\bibnamefont {Walsworth}},\ }\bibfield  {title} {\bibinfo
  {title} {Solid-state electronic spin coherence time approaching one second},\
  }\href {https://doi.org/10.1038/ncomms2771} {\bibfield  {journal} {\bibinfo
  {journal} {Nat. Commun.}\ }\textbf {\bibinfo {volume} {4}},\ \bibinfo {pages}
  {1743} (\bibinfo {year} {2013})}\BibitemShut {NoStop}%
\bibitem [{\citenamefont {Bar-Gill}\ \emph {et~al.}(2009)\citenamefont
  {Bar-Gill}, \citenamefont {Pham}, \citenamefont {Jarmola}, \citenamefont
  {Budker},\ and\ \citenamefont {Walsworth}}]{Balasubramanian2009}%
  \BibitemOpen
  \bibfield  {author} {\bibinfo {author} {\bibfnamefont {N.}~\bibnamefont
  {Bar-Gill}}, \bibinfo {author} {\bibfnamefont {L.}~\bibnamefont {Pham}},
  \bibinfo {author} {\bibfnamefont {A.}~\bibnamefont {Jarmola}}, \bibinfo
  {author} {\bibfnamefont {D.}~\bibnamefont {Budker}},\ and\ \bibinfo {author}
  {\bibfnamefont {R.}~\bibnamefont {Walsworth}},\ }\bibfield  {title} {\bibinfo
  {title} {Ultralong spin coherence time in isotopically engineered diamond},\
  }\href {https://doi.org/10.1038/nmat2420} {\bibfield  {journal} {\bibinfo
  {journal} {Nat. Mater.}\ }\textbf {\bibinfo {volume} {8}},\ \bibinfo {pages}
  {383} (\bibinfo {year} {2009})}\BibitemShut {NoStop}%
\bibitem [{\citenamefont {Maurer}\ \emph {et~al.}(2012)\citenamefont {Maurer},
  \citenamefont {Kucsko}, \citenamefont {Latta}, \citenamefont {Jiang},
  \citenamefont {Yao}, \citenamefont {Bennett}, \citenamefont {Pastawski},
  \citenamefont {Hunger}, \citenamefont {Chisholm}, \citenamefont {Markham},
  \citenamefont {Twitchen}, \citenamefont {Cirac},\ and\ \citenamefont
  {Lukin}}]{Maurer1283}%
  \BibitemOpen
  \bibfield  {author} {\bibinfo {author} {\bibfnamefont {P.~C.}\ \bibnamefont
  {Maurer}}, \bibinfo {author} {\bibfnamefont {G.}~\bibnamefont {Kucsko}},
  \bibinfo {author} {\bibfnamefont {C.}~\bibnamefont {Latta}}, \bibinfo
  {author} {\bibfnamefont {L.}~\bibnamefont {Jiang}}, \bibinfo {author}
  {\bibfnamefont {N.~Y.}\ \bibnamefont {Yao}}, \bibinfo {author} {\bibfnamefont
  {S.~D.}\ \bibnamefont {Bennett}}, \bibinfo {author} {\bibfnamefont
  {F.}~\bibnamefont {Pastawski}}, \bibinfo {author} {\bibfnamefont
  {D.}~\bibnamefont {Hunger}}, \bibinfo {author} {\bibfnamefont
  {N.}~\bibnamefont {Chisholm}}, \bibinfo {author} {\bibfnamefont
  {M.}~\bibnamefont {Markham}}, \bibinfo {author} {\bibfnamefont {D.~J.}\
  \bibnamefont {Twitchen}}, \bibinfo {author} {\bibfnamefont {J.~I.}\
  \bibnamefont {Cirac}},\ and\ \bibinfo {author} {\bibfnamefont {M.~D.}\
  \bibnamefont {Lukin}},\ }\bibfield  {title} {\bibinfo {title}
  {Room-temperature quantum bit memory exceeding one second},\ }\href
  {https://doi.org/10.1126/science.1220513} {\bibfield  {journal} {\bibinfo
  {journal} {Science}\ }\textbf {\bibinfo {volume} {336}},\ \bibinfo {pages}
  {1283} (\bibinfo {year} {2012})}\BibitemShut {NoStop}%
\bibitem [{\citenamefont {Kennedy}\ \emph {et~al.}(2003)\citenamefont
  {Kennedy}, \citenamefont {Colton}, \citenamefont {Butler}, \citenamefont
  {Linares},\ and\ \citenamefont {Doering}}]{doi:10.1063/1.1626791}%
  \BibitemOpen
  \bibfield  {author} {\bibinfo {author} {\bibfnamefont {T.~A.}\ \bibnamefont
  {Kennedy}}, \bibinfo {author} {\bibfnamefont {J.~S.}\ \bibnamefont {Colton}},
  \bibinfo {author} {\bibfnamefont {J.~E.}\ \bibnamefont {Butler}}, \bibinfo
  {author} {\bibfnamefont {R.~C.}\ \bibnamefont {Linares}},\ and\ \bibinfo
  {author} {\bibfnamefont {P.~J.}\ \bibnamefont {Doering}},\ }\bibfield
  {title} {\bibinfo {title} {Long coherence times at 300 k for nitrogen-vacancy
  center spins in diamond grown by chemical vapor deposition},\ }\href
  {https://doi.org/10.1063/1.1626791} {\bibfield  {journal} {\bibinfo
  {journal} {Appl. Phys. Lett.}\ }\textbf {\bibinfo {volume} {83}},\ \bibinfo
  {pages} {4190} (\bibinfo {year} {2003})}\BibitemShut {NoStop}%
\bibitem [{\citenamefont {Takahashi}\ \emph {et~al.}(2008)\citenamefont
  {Takahashi}, \citenamefont {Hanson}, \citenamefont {van Tol}, \citenamefont
  {Sherwin},\ and\ \citenamefont {Awschalom}}]{PhysRevLett.101.047601}%
  \BibitemOpen
  \bibfield  {author} {\bibinfo {author} {\bibfnamefont {S.}~\bibnamefont
  {Takahashi}}, \bibinfo {author} {\bibfnamefont {R.}~\bibnamefont {Hanson}},
  \bibinfo {author} {\bibfnamefont {J.}~\bibnamefont {van Tol}}, \bibinfo
  {author} {\bibfnamefont {M.~S.}\ \bibnamefont {Sherwin}},\ and\ \bibinfo
  {author} {\bibfnamefont {D.~D.}\ \bibnamefont {Awschalom}},\ }\bibfield
  {title} {\bibinfo {title} {Quenching spin decoherence in diamond through spin
  bath polarization},\ }\href {https://doi.org/10.1103/PhysRevLett.101.047601}
  {\bibfield  {journal} {\bibinfo  {journal} {Phys. Rev. Lett.}\ }\textbf
  {\bibinfo {volume} {101}},\ \bibinfo {pages} {047601} (\bibinfo {year}
  {2008})}\BibitemShut {NoStop}%
\bibitem [{\citenamefont {Du}\ \emph {et~al.}(2009)\citenamefont {Du},
  \citenamefont {Rong}, \citenamefont {Zhao}, \citenamefont {Wang},
  \citenamefont {Yang},\ and\ \citenamefont {Liu}}]{Du2009}%
  \BibitemOpen
  \bibfield  {author} {\bibinfo {author} {\bibfnamefont {J.}~\bibnamefont
  {Du}}, \bibinfo {author} {\bibfnamefont {X.}~\bibnamefont {Rong}}, \bibinfo
  {author} {\bibfnamefont {N.}~\bibnamefont {Zhao}}, \bibinfo {author}
  {\bibfnamefont {Y.}~\bibnamefont {Wang}}, \bibinfo {author} {\bibfnamefont
  {J.}~\bibnamefont {Yang}},\ and\ \bibinfo {author} {\bibfnamefont {R.~B.}\
  \bibnamefont {Liu}},\ }\bibfield  {title} {\bibinfo {title} {Preserving
  electron spin coherence in solids by optimal dynamical decoupling},\ }\href
  {https://doi.org/10.1038/nature08470} {\bibfield  {journal} {\bibinfo
  {journal} {Nature}\ }\textbf {\bibinfo {volume} {461}},\ \bibinfo {pages}
  {1265} (\bibinfo {year} {2009})}\BibitemShut {NoStop}%
\bibitem [{\citenamefont {Bassett}\ \emph {et~al.}(2014)\citenamefont
  {Bassett}, \citenamefont {Heremans}, \citenamefont {Christle}, \citenamefont
  {Yale}, \citenamefont {Burkard}, \citenamefont {Buckley},\ and\ \citenamefont
  {Awschalom}}]{Bassett1333}%
  \BibitemOpen
  \bibfield  {author} {\bibinfo {author} {\bibfnamefont {L.~C.}\ \bibnamefont
  {Bassett}}, \bibinfo {author} {\bibfnamefont {F.~J.}\ \bibnamefont
  {Heremans}}, \bibinfo {author} {\bibfnamefont {D.~J.}\ \bibnamefont
  {Christle}}, \bibinfo {author} {\bibfnamefont {C.~G.}\ \bibnamefont {Yale}},
  \bibinfo {author} {\bibfnamefont {G.}~\bibnamefont {Burkard}}, \bibinfo
  {author} {\bibfnamefont {B.~B.}\ \bibnamefont {Buckley}},\ and\ \bibinfo
  {author} {\bibfnamefont {D.~D.}\ \bibnamefont {Awschalom}},\ }\bibfield
  {title} {\bibinfo {title} {Ultrafast optical control of orbital and spin
  dynamics in a solid-state defect},\ }\href
  {https://doi.org/10.1126/science.1255541} {\bibfield  {journal} {\bibinfo
  {journal} {Science}\ }\textbf {\bibinfo {volume} {345}},\ \bibinfo {pages}
  {1333} (\bibinfo {year} {2014})}\BibitemShut {NoStop}%
\bibitem [{\citenamefont {Buckley}\ \emph {et~al.}(2010)\citenamefont
  {Buckley}, \citenamefont {Fuchs}, \citenamefont {Bassett},\ and\
  \citenamefont {Awschalom}}]{Buckley1212}%
  \BibitemOpen
  \bibfield  {author} {\bibinfo {author} {\bibfnamefont {B.~B.}\ \bibnamefont
  {Buckley}}, \bibinfo {author} {\bibfnamefont {G.~D.}\ \bibnamefont {Fuchs}},
  \bibinfo {author} {\bibfnamefont {L.~C.}\ \bibnamefont {Bassett}},\ and\
  \bibinfo {author} {\bibfnamefont {D.~D.}\ \bibnamefont {Awschalom}},\
  }\bibfield  {title} {\bibinfo {title} {Spin-light coherence for single-spin
  measurement and control in diamond},\ }\href
  {https://doi.org/10.1126/science.1196436} {\bibfield  {journal} {\bibinfo
  {journal} {Science}\ }\textbf {\bibinfo {volume} {330}},\ \bibinfo {pages}
  {1212} (\bibinfo {year} {2010})}\BibitemShut {NoStop}%
\bibitem [{\citenamefont {Steiner}\ \emph {et~al.}(2010)\citenamefont
  {Steiner}, \citenamefont {Neumann}, \citenamefont {Beck}, \citenamefont
  {Jelezko},\ and\ \citenamefont {Wrachtrup}}]{PhysRevB.81.035205}%
  \BibitemOpen
  \bibfield  {author} {\bibinfo {author} {\bibfnamefont {M.}~\bibnamefont
  {Steiner}}, \bibinfo {author} {\bibfnamefont {P.}~\bibnamefont {Neumann}},
  \bibinfo {author} {\bibfnamefont {J.}~\bibnamefont {Beck}}, \bibinfo {author}
  {\bibfnamefont {F.}~\bibnamefont {Jelezko}},\ and\ \bibinfo {author}
  {\bibfnamefont {J.}~\bibnamefont {Wrachtrup}},\ }\bibfield  {title} {\bibinfo
  {title} {Universal enhancement of the optical readout fidelity of single
  electron spins at nitrogen-vacancy centers in diamond},\ }\href
  {https://doi.org/10.1103/PhysRevB.81.035205} {\bibfield  {journal} {\bibinfo
  {journal} {Phys. Rev. B}\ }\textbf {\bibinfo {volume} {81}},\ \bibinfo
  {pages} {035205} (\bibinfo {year} {2010})}\BibitemShut {NoStop}%
\bibitem [{\citenamefont {Toyli}\ \emph {et~al.}(2012)\citenamefont {Toyli},
  \citenamefont {Christle}, \citenamefont {Alkauskas}, \citenamefont {Buckley},
  \citenamefont {Van~de Walle},\ and\ \citenamefont
  {Awschalom}}]{PhysRevX.2.031001}%
  \BibitemOpen
  \bibfield  {author} {\bibinfo {author} {\bibfnamefont {D.~M.}\ \bibnamefont
  {Toyli}}, \bibinfo {author} {\bibfnamefont {D.~J.}\ \bibnamefont {Christle}},
  \bibinfo {author} {\bibfnamefont {A.}~\bibnamefont {Alkauskas}}, \bibinfo
  {author} {\bibfnamefont {B.~B.}\ \bibnamefont {Buckley}}, \bibinfo {author}
  {\bibfnamefont {C.~G.}\ \bibnamefont {Van~de Walle}},\ and\ \bibinfo {author}
  {\bibfnamefont {D.~D.}\ \bibnamefont {Awschalom}},\ }\bibfield  {title}
  {\bibinfo {title} {Measurement and control of single nitrogen-vacancy center
  spins above 600 k},\ }\href {https://doi.org/10.1103/PhysRevX.2.031001}
  {\bibfield  {journal} {\bibinfo  {journal} {Phys. Rev. X}\ }\textbf {\bibinfo
  {volume} {2}},\ \bibinfo {pages} {031001} (\bibinfo {year}
  {2012})}\BibitemShut {NoStop}%
\bibitem [{\citenamefont {Tamarat}\ \emph {et~al.}(2006)\citenamefont
  {Tamarat}, \citenamefont {Gaebel}, \citenamefont {Rabeau}, \citenamefont
  {Khan}, \citenamefont {Greentree}, \citenamefont {Wilson}, \citenamefont
  {Hollenberg}, \citenamefont {Prawer}, \citenamefont {Hemmer}, \citenamefont
  {Jelezko},\ and\ \citenamefont {Wrachtrup}}]{PhysRevLett.97.083002}%
  \BibitemOpen
  \bibfield  {author} {\bibinfo {author} {\bibfnamefont {P.}~\bibnamefont
  {Tamarat}}, \bibinfo {author} {\bibfnamefont {T.}~\bibnamefont {Gaebel}},
  \bibinfo {author} {\bibfnamefont {J.~R.}\ \bibnamefont {Rabeau}}, \bibinfo
  {author} {\bibfnamefont {M.}~\bibnamefont {Khan}}, \bibinfo {author}
  {\bibfnamefont {A.~D.}\ \bibnamefont {Greentree}}, \bibinfo {author}
  {\bibfnamefont {H.}~\bibnamefont {Wilson}}, \bibinfo {author} {\bibfnamefont
  {L.~C.~L.}\ \bibnamefont {Hollenberg}}, \bibinfo {author} {\bibfnamefont
  {S.}~\bibnamefont {Prawer}}, \bibinfo {author} {\bibfnamefont
  {P.}~\bibnamefont {Hemmer}}, \bibinfo {author} {\bibfnamefont
  {F.}~\bibnamefont {Jelezko}},\ and\ \bibinfo {author} {\bibfnamefont
  {J.}~\bibnamefont {Wrachtrup}},\ }\bibfield  {title} {\bibinfo {title} {Stark
  shift control of single optical centers in diamond},\ }\href
  {https://doi.org/10.1103/PhysRevLett.97.083002} {\bibfield  {journal}
  {\bibinfo  {journal} {Phys. Rev. Lett.}\ }\textbf {\bibinfo {volume} {97}},\
  \bibinfo {pages} {083002} (\bibinfo {year} {2006})}\BibitemShut {NoStop}%
\bibitem [{\citenamefont {Siyushev}\ \emph {et~al.}(2013)\citenamefont
  {Siyushev}, \citenamefont {Pinto}, \citenamefont {V\"or\"os}, \citenamefont
  {Gali}, \citenamefont {Jelezko},\ and\ \citenamefont
  {Wrachtrup}}]{PhysRevLett.110.167402}%
  \BibitemOpen
  \bibfield  {author} {\bibinfo {author} {\bibfnamefont {P.}~\bibnamefont
  {Siyushev}}, \bibinfo {author} {\bibfnamefont {H.}~\bibnamefont {Pinto}},
  \bibinfo {author} {\bibfnamefont {M.}~\bibnamefont {V\"or\"os}}, \bibinfo
  {author} {\bibfnamefont {A.}~\bibnamefont {Gali}}, \bibinfo {author}
  {\bibfnamefont {F.}~\bibnamefont {Jelezko}},\ and\ \bibinfo {author}
  {\bibfnamefont {J.}~\bibnamefont {Wrachtrup}},\ }\bibfield  {title} {\bibinfo
  {title} {Optically controlled switching of the charge state of a single
  nitrogen-vacancy center in diamond at cryogenic temperatures},\ }\href
  {https://doi.org/10.1103/PhysRevLett.110.167402} {\bibfield  {journal}
  {\bibinfo  {journal} {Phys. Rev. Lett.}\ }\textbf {\bibinfo {volume} {110}},\
  \bibinfo {pages} {167402} (\bibinfo {year} {2013})}\BibitemShut {NoStop}%
\bibitem [{\citenamefont {Gao}\ \emph {et~al.}(2015)\citenamefont {Gao},
  \citenamefont {Imamoglu}, \citenamefont {Bernien},\ and\ \citenamefont
  {Hanson}}]{Gao2015}%
  \BibitemOpen
  \bibfield  {author} {\bibinfo {author} {\bibfnamefont {W.~B.}\ \bibnamefont
  {Gao}}, \bibinfo {author} {\bibfnamefont {A.}~\bibnamefont {Imamoglu}},
  \bibinfo {author} {\bibfnamefont {H.}~\bibnamefont {Bernien}},\ and\ \bibinfo
  {author} {\bibfnamefont {R.}~\bibnamefont {Hanson}},\ }\bibfield  {title}
  {\bibinfo {title} {Coherent manipulation, measurement and entanglement of
  individual solid-state spins using optical fields},\ }\href
  {https://doi.org/10.1038/nphoton.2015.58} {\bibfield  {journal} {\bibinfo
  {journal} {Nat. Photonics}\ }\textbf {\bibinfo {volume} {9}},\ \bibinfo
  {pages} {363} (\bibinfo {year} {2015})}\BibitemShut {NoStop}%
\bibitem [{\citenamefont {Li}\ \emph {et~al.}(2020{\natexlab{a}})\citenamefont
  {Li}, \citenamefont {Zhou}, \citenamefont {Gao},\ and\ \citenamefont
  {Nori}}]{PhysRevLett.125.153602}%
  \BibitemOpen
  \bibfield  {author} {\bibinfo {author} {\bibfnamefont {P.-B.}\ \bibnamefont
  {Li}}, \bibinfo {author} {\bibfnamefont {Y.}~\bibnamefont {Zhou}}, \bibinfo
  {author} {\bibfnamefont {W.-B.}\ \bibnamefont {Gao}},\ and\ \bibinfo {author}
  {\bibfnamefont {F.}~\bibnamefont {Nori}},\ }\bibfield  {title} {\bibinfo
  {title} {Enhancing spin-phonon and spin-spin interactions using linear
  resources in a hybrid quantum system},\ }\href
  {https://doi.org/10.1103/PhysRevLett.125.153602} {\bibfield  {journal}
  {\bibinfo  {journal} {Phys. Rev. Lett.}\ }\textbf {\bibinfo {volume} {125}},\
  \bibinfo {pages} {153602} (\bibinfo {year} {2020}{\natexlab{a}})}\BibitemShut
  {NoStop}%
\bibitem [{\citenamefont {Rabl}\ \emph {et~al.}(2009)\citenamefont {Rabl},
  \citenamefont {Cappellaro}, \citenamefont {Dutt}, \citenamefont {Jiang},
  \citenamefont {Maze},\ and\ \citenamefont {Lukin}}]{PhysRevB.79.041302}%
  \BibitemOpen
  \bibfield  {author} {\bibinfo {author} {\bibfnamefont {P.}~\bibnamefont
  {Rabl}}, \bibinfo {author} {\bibfnamefont {P.}~\bibnamefont {Cappellaro}},
  \bibinfo {author} {\bibfnamefont {M.~V.~G.}\ \bibnamefont {Dutt}}, \bibinfo
  {author} {\bibfnamefont {L.}~\bibnamefont {Jiang}}, \bibinfo {author}
  {\bibfnamefont {J.~R.}\ \bibnamefont {Maze}},\ and\ \bibinfo {author}
  {\bibfnamefont {M.~D.}\ \bibnamefont {Lukin}},\ }\bibfield  {title} {\bibinfo
  {title} {Strong magnetic coupling between an electronic spin qubit and a
  mechanical resonator},\ }\href {https://doi.org/10.1103/PhysRevB.79.041302}
  {\bibfield  {journal} {\bibinfo  {journal} {Phys. Rev. B}\ }\textbf {\bibinfo
  {volume} {79}},\ \bibinfo {pages} {041302} (\bibinfo {year}
  {2009})}\BibitemShut {NoStop}%
\bibitem [{\citenamefont {Zhou}\ \emph {et~al.}(2018)\citenamefont {Zhou},
  \citenamefont {Li}, \citenamefont {Li}, \citenamefont {Li},\ and\
  \citenamefont {Li}}]{PhysRevA.98.052346}%
  \BibitemOpen
  \bibfield  {author} {\bibinfo {author} {\bibfnamefont {Y.}~\bibnamefont
  {Zhou}}, \bibinfo {author} {\bibfnamefont {B.}~\bibnamefont {Li}}, \bibinfo
  {author} {\bibfnamefont {X.-X.}\ \bibnamefont {Li}}, \bibinfo {author}
  {\bibfnamefont {F.-L.}\ \bibnamefont {Li}},\ and\ \bibinfo {author}
  {\bibfnamefont {P.-B.}\ \bibnamefont {Li}},\ }\bibfield  {title} {\bibinfo
  {title} {Preparing multiparticle entangled states of nitrogen-vacancy centers
  via adiabatic ground-state transitions},\ }\href
  {https://doi.org/10.1103/PhysRevA.98.052346} {\bibfield  {journal} {\bibinfo
  {journal} {Phys. Rev. A}\ }\textbf {\bibinfo {volume} {98}},\ \bibinfo
  {pages} {052346} (\bibinfo {year} {2018})}\BibitemShut {NoStop}%
\bibitem [{\citenamefont {Arcizet}\ \emph {et~al.}(2011)\citenamefont
  {Arcizet}, \citenamefont {Jacques}, \citenamefont {Siria}, \citenamefont
  {Poncharal}, \citenamefont {Vincent},\ and\ \citenamefont
  {Seidelin}}]{Arcizet2011}%
  \BibitemOpen
  \bibfield  {author} {\bibinfo {author} {\bibfnamefont {O.}~\bibnamefont
  {Arcizet}}, \bibinfo {author} {\bibfnamefont {V.}~\bibnamefont {Jacques}},
  \bibinfo {author} {\bibfnamefont {A.}~\bibnamefont {Siria}}, \bibinfo
  {author} {\bibfnamefont {P.}~\bibnamefont {Poncharal}}, \bibinfo {author}
  {\bibfnamefont {P.}~\bibnamefont {Vincent}},\ and\ \bibinfo {author}
  {\bibfnamefont {S.}~\bibnamefont {Seidelin}},\ }\bibfield  {title} {\bibinfo
  {title} {A single nitrogen-vacancy defect coupled to a nanomechanical
  oscillator},\ }\href {https://doi.org/10.1038/nphys2070} {\bibfield
  {journal} {\bibinfo  {journal} {Nat. Phys.}\ }\textbf {\bibinfo {volume}
  {7}},\ \bibinfo {pages} {879} (\bibinfo {year} {2011})}\BibitemShut {NoStop}%
\bibitem [{\citenamefont {Bennett}\ \emph {et~al.}(2013)\citenamefont
  {Bennett}, \citenamefont {Yao}, \citenamefont {Otterbach}, \citenamefont
  {Zoller}, \citenamefont {Rabl},\ and\ \citenamefont
  {Lukin}}]{PhysRevLett.110.156402}%
  \BibitemOpen
  \bibfield  {author} {\bibinfo {author} {\bibfnamefont {S.~D.}\ \bibnamefont
  {Bennett}}, \bibinfo {author} {\bibfnamefont {N.~Y.}\ \bibnamefont {Yao}},
  \bibinfo {author} {\bibfnamefont {J.}~\bibnamefont {Otterbach}}, \bibinfo
  {author} {\bibfnamefont {P.}~\bibnamefont {Zoller}}, \bibinfo {author}
  {\bibfnamefont {P.}~\bibnamefont {Rabl}},\ and\ \bibinfo {author}
  {\bibfnamefont {M.~D.}\ \bibnamefont {Lukin}},\ }\bibfield  {title} {\bibinfo
  {title} {Phonon-induced spin-spin interactions in diamond nanostructures:
  Application to spin squeezing},\ }\href
  {https://doi.org/10.1103/PhysRevLett.110.156402} {\bibfield  {journal}
  {\bibinfo  {journal} {Phys. Rev. Lett.}\ }\textbf {\bibinfo {volume} {110}},\
  \bibinfo {pages} {156402} (\bibinfo {year} {2013})}\BibitemShut {NoStop}%
\bibitem [{\citenamefont {Teissier}\ \emph {et~al.}(2014)\citenamefont
  {Teissier}, \citenamefont {Barfuss}, \citenamefont {Appel}, \citenamefont
  {Neu},\ and\ \citenamefont {Maletinsky}}]{PhysRevLett.113.020503}%
  \BibitemOpen
  \bibfield  {author} {\bibinfo {author} {\bibfnamefont {J.}~\bibnamefont
  {Teissier}}, \bibinfo {author} {\bibfnamefont {A.}~\bibnamefont {Barfuss}},
  \bibinfo {author} {\bibfnamefont {P.}~\bibnamefont {Appel}}, \bibinfo
  {author} {\bibfnamefont {E.}~\bibnamefont {Neu}},\ and\ \bibinfo {author}
  {\bibfnamefont {P.}~\bibnamefont {Maletinsky}},\ }\bibfield  {title}
  {\bibinfo {title} {Strain coupling of a nitrogen-vacancy center spin to a
  diamond mechanical oscillator},\ }\href
  {https://doi.org/10.1103/PhysRevLett.113.020503} {\bibfield  {journal}
  {\bibinfo  {journal} {Phys. Rev. Lett.}\ }\textbf {\bibinfo {volume} {113}},\
  \bibinfo {pages} {020503} (\bibinfo {year} {2014})}\BibitemShut {NoStop}%
\bibitem [{\citenamefont {Ovartchaiyapong}\ \emph {et~al.}(2014)\citenamefont
  {Ovartchaiyapong}, \citenamefont {Lee}, \citenamefont {Myers},\ and\
  \citenamefont {Jayich}}]{Ovartchaiyapong2014}%
  \BibitemOpen
  \bibfield  {author} {\bibinfo {author} {\bibfnamefont {P.}~\bibnamefont
  {Ovartchaiyapong}}, \bibinfo {author} {\bibfnamefont {K.~W.}\ \bibnamefont
  {Lee}}, \bibinfo {author} {\bibfnamefont {B.~A.}\ \bibnamefont {Myers}},\
  and\ \bibinfo {author} {\bibfnamefont {A.~C.~B.}\ \bibnamefont {Jayich}},\
  }\bibfield  {title} {\bibinfo {title} {Dynamic strain-mediated coupling of a
  single diamond spin to a mechanical resonator},\ }\href
  {https://doi.org/10.1038/ncomms5429} {\bibfield  {journal} {\bibinfo
  {journal} {Nat. Commun.}\ }\textbf {\bibinfo {volume} {5}},\ \bibinfo {pages}
  {4429} (\bibinfo {year} {2014})}\BibitemShut {NoStop}%
\bibitem [{\citenamefont {Li}\ \emph {et~al.}(2020{\natexlab{b}})\citenamefont
  {Li}, \citenamefont {Li},\ and\ \citenamefont {Li}}]{li2020simulation}%
  \BibitemOpen
  \bibfield  {author} {\bibinfo {author} {\bibfnamefont {X.-X.}\ \bibnamefont
  {Li}}, \bibinfo {author} {\bibfnamefont {B.}~\bibnamefont {Li}},\ and\
  \bibinfo {author} {\bibfnamefont {P.-B.}\ \bibnamefont {Li}},\ }\bibfield
  {title} {\bibinfo {title} {Simulation of topological phases with color center
  arrays in phononic crystals},\ }\href {https://doi.org/0} {\bibfield
  {journal} {\bibinfo  {journal} {Phys. Rev. Research}\ }\textbf {\bibinfo
  {volume} {2}},\ \bibinfo {pages} {013121} (\bibinfo {year}
  {2020}{\natexlab{b}})}\BibitemShut {NoStop}%
\bibitem [{\citenamefont {Li}\ \emph {et~al.}(2021)\citenamefont {Li},
  \citenamefont {Li}, \citenamefont {Li}, \citenamefont {Gao},\ and\
  \citenamefont {Li}}]{li2021simulation}%
  \BibitemOpen
  \bibfield  {author} {\bibinfo {author} {\bibfnamefont {X.-X.}\ \bibnamefont
  {Li}}, \bibinfo {author} {\bibfnamefont {P.-B.}\ \bibnamefont {Li}}, \bibinfo
  {author} {\bibfnamefont {H.-R.}\ \bibnamefont {Li}}, \bibinfo {author}
  {\bibfnamefont {H.}~\bibnamefont {Gao}},\ and\ \bibinfo {author}
  {\bibfnamefont {F.-L.}\ \bibnamefont {Li}},\ }\bibfield  {title} {\bibinfo
  {title} {Simulation of topological zak phase in spin-phononic crystal
  networks},\ }\href {https://doi.org/0} {\bibfield  {journal} {\bibinfo
  {journal} {Phys. Rev. Research}\ }\textbf {\bibinfo {volume} {3}},\ \bibinfo
  {pages} {013025} (\bibinfo {year} {2021})}\BibitemShut {NoStop}%
\bibitem [{\citenamefont {Dong}\ \emph {et~al.}(2021)\citenamefont {Dong},
  \citenamefont {Li}, \citenamefont {Liu},\ and\ \citenamefont
  {Nori}}]{dong2021unconventional}%
  \BibitemOpen
  \bibfield  {author} {\bibinfo {author} {\bibfnamefont {X.-L.}\ \bibnamefont
  {Dong}}, \bibinfo {author} {\bibfnamefont {P.-B.}\ \bibnamefont {Li}},
  \bibinfo {author} {\bibfnamefont {T.}~\bibnamefont {Liu}},\ and\ \bibinfo
  {author} {\bibfnamefont {F.}~\bibnamefont {Nori}},\ }\bibfield  {title}
  {\bibinfo {title} {Unconventional quantum sound-matter interactions in
  spin-optomechanical-crystal hybrid systems},\ }\href {https://doi.org/0}
  {\bibfield  {journal} {\bibinfo  {journal} {Phys. Rev. Lett.}\ }\textbf
  {\bibinfo {volume} {126}},\ \bibinfo {pages} {203601} (\bibinfo {year}
  {2021})}\BibitemShut {NoStop}%
\bibitem [{\citenamefont {Winkler}\ \emph {et~al.}(2006)\citenamefont
  {Winkler}, \citenamefont {Thalhammer}, \citenamefont {Lang}, \citenamefont
  {Grimm}, \citenamefont {Hecker~Denschlag}, \citenamefont {Daley},
  \citenamefont {Kantian}, \citenamefont {B\"{u}chler},\ and\ \citenamefont
  {Zoller}}]{Winkler2006}%
  \BibitemOpen
  \bibfield  {author} {\bibinfo {author} {\bibfnamefont {K.}~\bibnamefont
  {Winkler}}, \bibinfo {author} {\bibfnamefont {G.}~\bibnamefont {Thalhammer}},
  \bibinfo {author} {\bibfnamefont {F.}~\bibnamefont {Lang}}, \bibinfo {author}
  {\bibfnamefont {R.}~\bibnamefont {Grimm}}, \bibinfo {author} {\bibfnamefont
  {J.}~\bibnamefont {Hecker~Denschlag}}, \bibinfo {author} {\bibfnamefont
  {A.~J.}\ \bibnamefont {Daley}}, \bibinfo {author} {\bibfnamefont
  {A.}~\bibnamefont {Kantian}}, \bibinfo {author} {\bibfnamefont {H.~P.}\
  \bibnamefont {B\"{u}chler}},\ and\ \bibinfo {author} {\bibfnamefont
  {P.}~\bibnamefont {Zoller}},\ }\bibfield  {title} {\bibinfo {title}
  {Repulsively bound atom pairs in an optical lattice},\ }\href
  {https://doi.org/10.1038/nature04918} {\bibfield  {journal} {\bibinfo
  {journal} {Nature}\ }\textbf {\bibinfo {volume} {441}},\ \bibinfo {pages}
  {853} (\bibinfo {year} {2006})}\BibitemShut {NoStop}%
\bibitem [{\citenamefont {L{\"u}}\ \emph {et~al.}(2015)\citenamefont {L{\"u}},
  \citenamefont {Liao}, \citenamefont {Tian},\ and\ \citenamefont
  {Nori}}]{lu2015steady}%
  \BibitemOpen
  \bibfield  {author} {\bibinfo {author} {\bibfnamefont {X.-Y.}\ \bibnamefont
  {L{\"u}}}, \bibinfo {author} {\bibfnamefont {J.-Q.}\ \bibnamefont {Liao}},
  \bibinfo {author} {\bibfnamefont {L.}~\bibnamefont {Tian}},\ and\ \bibinfo
  {author} {\bibfnamefont {F.}~\bibnamefont {Nori}},\ }\bibfield  {title}
  {\bibinfo {title} {Steady-state mechanical squeezing in an optomechanical
  system via {Duffing} nonlinearity},\ }\href {https://doi.org/0} {\bibfield
  {journal} {\bibinfo  {journal} {Phys. Rev. A}\ }\textbf {\bibinfo {volume}
  {91}},\ \bibinfo {pages} {013834} (\bibinfo {year} {2015})}\BibitemShut
  {NoStop}%
\bibitem [{\citenamefont {L\"{u}}\ \emph {et~al.}(2013)\citenamefont {L\"{u}},
  \citenamefont {Zhang}, \citenamefont {Ashhab}, \citenamefont {Wu},\ and\
  \citenamefont {Nori}}]{lu2013quantum}%
  \BibitemOpen
  \bibfield  {author} {\bibinfo {author} {\bibfnamefont {X.-Y.}\ \bibnamefont
  {L\"{u}}}, \bibinfo {author} {\bibfnamefont {W.-M.}\ \bibnamefont {Zhang}},
  \bibinfo {author} {\bibfnamefont {S.}~\bibnamefont {Ashhab}}, \bibinfo
  {author} {\bibfnamefont {Y.}~\bibnamefont {Wu}},\ and\ \bibinfo {author}
  {\bibfnamefont {F.}~\bibnamefont {Nori}},\ }\bibfield  {title} {\bibinfo
  {title} {Quantum-criticality-induced strong kerr nonlinearities in
  optomechanical systems},\ }\href {https://doi.org/10.1038/srep02943}
  {\bibfield  {journal} {\bibinfo  {journal} {Sci. Rep.}\ }\textbf {\bibinfo
  {volume} {3}},\ \bibinfo {pages} {1} (\bibinfo {year} {2013})}\BibitemShut
  {NoStop}%
\bibitem [{\citenamefont {Rips}\ \emph {et~al.}(2012)\citenamefont {Rips},
  \citenamefont {Kiffner}, \citenamefont {Wilson-Rae},\ and\ \citenamefont
  {Hartmann}}]{Rips_2012}%
  \BibitemOpen
  \bibfield  {author} {\bibinfo {author} {\bibfnamefont {S.}~\bibnamefont
  {Rips}}, \bibinfo {author} {\bibfnamefont {M.}~\bibnamefont {Kiffner}},
  \bibinfo {author} {\bibfnamefont {I.}~\bibnamefont {Wilson-Rae}},\ and\
  \bibinfo {author} {\bibfnamefont {M.~J.}\ \bibnamefont {Hartmann}},\
  }\bibfield  {title} {\bibinfo {title} {Steady-state negative wigner functions
  of nonlinear nanomechanical oscillators},\ }\href
  {https://doi.org/10.1088/1367-2630/14/2/023042} {\bibfield  {journal}
  {\bibinfo  {journal} {New J. Phys.}\ }\textbf {\bibinfo {volume} {14}},\
  \bibinfo {pages} {023042} (\bibinfo {year} {2012})}\BibitemShut {NoStop}%
\bibitem [{\citenamefont {Doherty}\ \emph {et~al.}(2012)\citenamefont
  {Doherty}, \citenamefont {Dolde}, \citenamefont {Fedder}, \citenamefont
  {Jelezko}, \citenamefont {Wrachtrup}, \citenamefont {Manson},\ and\
  \citenamefont {Hollenberg}}]{PhysRevB.85.205203}%
  \BibitemOpen
  \bibfield  {author} {\bibinfo {author} {\bibfnamefont {M.~W.}\ \bibnamefont
  {Doherty}}, \bibinfo {author} {\bibfnamefont {F.}~\bibnamefont {Dolde}},
  \bibinfo {author} {\bibfnamefont {H.}~\bibnamefont {Fedder}}, \bibinfo
  {author} {\bibfnamefont {F.}~\bibnamefont {Jelezko}}, \bibinfo {author}
  {\bibfnamefont {J.}~\bibnamefont {Wrachtrup}}, \bibinfo {author}
  {\bibfnamefont {N.~B.}\ \bibnamefont {Manson}},\ and\ \bibinfo {author}
  {\bibfnamefont {L.~C.~L.}\ \bibnamefont {Hollenberg}},\ }\bibfield  {title}
  {\bibinfo {title} {Theory of the ground-state spin of the
  \text{NV}${}^{\ensuremath{-}}$ center in diamond},\ }\href
  {https://doi.org/10.1103/PhysRevB.85.205203} {\bibfield  {journal} {\bibinfo
  {journal} {Phys. Rev. B}\ }\textbf {\bibinfo {volume} {85}},\ \bibinfo
  {pages} {205203} (\bibinfo {year} {2012})}\BibitemShut {NoStop}%
\bibitem [{\citenamefont {Doherty}\ \emph {et~al.}(2013)\citenamefont
  {Doherty}, \citenamefont {Manson}, \citenamefont {Delaney}, \citenamefont
  {Jelezko}, \citenamefont {Wrachtrup},\ and\ \citenamefont
  {Hollenberg}}]{DOHERTY20131}%
  \BibitemOpen
  \bibfield  {author} {\bibinfo {author} {\bibfnamefont {M.~W.}\ \bibnamefont
  {Doherty}}, \bibinfo {author} {\bibfnamefont {N.~B.}\ \bibnamefont {Manson}},
  \bibinfo {author} {\bibfnamefont {P.}~\bibnamefont {Delaney}}, \bibinfo
  {author} {\bibfnamefont {F.}~\bibnamefont {Jelezko}}, \bibinfo {author}
  {\bibfnamefont {J.}~\bibnamefont {Wrachtrup}},\ and\ \bibinfo {author}
  {\bibfnamefont {L.~C.}\ \bibnamefont {Hollenberg}},\ }\bibfield  {title}
  {\bibinfo {title} {The nitrogen-vacancy colour centre in diamond},\ }\href
  {https://doi.org/https://doi.org/10.1016/j.physrep.2013.02.001} {\bibfield
  {journal} {\bibinfo  {journal} {Phys. Rep.}\ }\textbf {\bibinfo {volume}
  {528}},\ \bibinfo {pages} {1} (\bibinfo {year} {2013})}\BibitemShut {NoStop}%
\bibitem [{\citenamefont {Lee}\ \emph {et~al.}(2017)\citenamefont {Lee},
  \citenamefont {Lee}, \citenamefont {Cady}, \citenamefont {Ovartchaiyapong},\
  and\ \citenamefont {Jayich}}]{Lee_2017}%
  \BibitemOpen
  \bibfield  {author} {\bibinfo {author} {\bibfnamefont {D.}~\bibnamefont
  {Lee}}, \bibinfo {author} {\bibfnamefont {K.~W.}\ \bibnamefont {Lee}},
  \bibinfo {author} {\bibfnamefont {J.~V.}\ \bibnamefont {Cady}}, \bibinfo
  {author} {\bibfnamefont {P.}~\bibnamefont {Ovartchaiyapong}},\ and\ \bibinfo
  {author} {\bibfnamefont {A.~C.~B.}\ \bibnamefont {Jayich}},\ }\bibfield
  {title} {\bibinfo {title} {Topical review: spins and mechanics in diamond},\
  }\href {https://doi.org/10.1088/2040-8986/aa52cd} {\bibfield  {journal}
  {\bibinfo  {journal} {J. Opt.}\ }\textbf {\bibinfo {volume} {19}},\ \bibinfo
  {pages} {033001} (\bibinfo {year} {2017})}\BibitemShut {NoStop}%
\bibitem [{\citenamefont {Calaj\'o}\ \emph {et~al.}(2016)\citenamefont
  {Calaj\'o}, \citenamefont {Ciccarello}, \citenamefont {Chang},\ and\
  \citenamefont {Rabl}}]{PhysRevA.93.033833}%
  \BibitemOpen
  \bibfield  {author} {\bibinfo {author} {\bibfnamefont {G.}~\bibnamefont
  {Calaj\'o}}, \bibinfo {author} {\bibfnamefont {F.}~\bibnamefont
  {Ciccarello}}, \bibinfo {author} {\bibfnamefont {D.}~\bibnamefont {Chang}},\
  and\ \bibinfo {author} {\bibfnamefont {P.}~\bibnamefont {Rabl}},\ }\bibfield
  {title} {\bibinfo {title} {Atom-field dressed states in slow-light waveguide
  \text{QED}},\ }\href {https://doi.org/10.1103/PhysRevA.93.033833} {\bibfield
  {journal} {\bibinfo  {journal} {Phys. Rev. A}\ }\textbf {\bibinfo {volume}
  {93}},\ \bibinfo {pages} {033833} (\bibinfo {year} {2016})}\BibitemShut
  {NoStop}%
\bibitem [{\citenamefont {Valiente}\ and\ \citenamefont
  {Petrosyan}(2008)}]{Valiente_2008}%
  \BibitemOpen
  \bibfield  {author} {\bibinfo {author} {\bibfnamefont {M.}~\bibnamefont
  {Valiente}}\ and\ \bibinfo {author} {\bibfnamefont {D.}~\bibnamefont
  {Petrosyan}},\ }\bibfield  {title} {\bibinfo {title} {Two-particle states in
  the hubbard model},\ }\href {https://doi.org/10.1088/0953-4075/41/16/161002}
  {\bibfield  {journal} {\bibinfo  {journal} {J. Phys. B: At. Mol. Opt. Phys.}\
  }\textbf {\bibinfo {volume} {41}},\ \bibinfo {pages} {161002} (\bibinfo
  {year} {2008})}\BibitemShut {NoStop}%
\bibitem [{\citenamefont {Wang}\ \emph {et~al.}(2020)\citenamefont {Wang},
  \citenamefont {Jaako}, \citenamefont {Kirton},\ and\ \citenamefont
  {Rabl}}]{PhysRevLett.124.213601}%
  \BibitemOpen
  \bibfield  {author} {\bibinfo {author} {\bibfnamefont {Z.}~\bibnamefont
  {Wang}}, \bibinfo {author} {\bibfnamefont {T.}~\bibnamefont {Jaako}},
  \bibinfo {author} {\bibfnamefont {P.}~\bibnamefont {Kirton}},\ and\ \bibinfo
  {author} {\bibfnamefont {P.}~\bibnamefont {Rabl}},\ }\bibfield  {title}
  {\bibinfo {title} {Supercorrelated radiance in nonlinear photonic
  waveguides},\ }\href {https://doi.org/10.1103/PhysRevLett.124.213601}
  {\bibfield  {journal} {\bibinfo  {journal} {Phys. Rev. Lett.}\ }\textbf
  {\bibinfo {volume} {124}},\ \bibinfo {pages} {213601} (\bibinfo {year}
  {2020})}\BibitemShut {NoStop}%
\bibitem [{\citenamefont {John}\ and\ \citenamefont
  {Wang}(1990)}]{PhysRevLett.64.2418}%
  \BibitemOpen
  \bibfield  {author} {\bibinfo {author} {\bibfnamefont {S.}~\bibnamefont
  {John}}\ and\ \bibinfo {author} {\bibfnamefont {J.}~\bibnamefont {Wang}},\
  }\bibfield  {title} {\bibinfo {title} {Quantum electrodynamics near a
  photonic band gap: Photon bound states and dressed atoms},\ }\href
  {https://doi.org/10.1103/PhysRevLett.64.2418} {\bibfield  {journal} {\bibinfo
   {journal} {Phys. Rev. Lett.}\ }\textbf {\bibinfo {volume} {64}},\ \bibinfo
  {pages} {2418} (\bibinfo {year} {1990})}\BibitemShut {NoStop}%
\bibitem [{\citenamefont {John}\ and\ \citenamefont
  {Wang}(1991)}]{PhysRevB.43.12772}%
  \BibitemOpen
  \bibfield  {author} {\bibinfo {author} {\bibfnamefont {S.}~\bibnamefont
  {John}}\ and\ \bibinfo {author} {\bibfnamefont {J.}~\bibnamefont {Wang}},\
  }\bibfield  {title} {\bibinfo {title} {Quantum optics of localized light in a
  photonic band gap},\ }\href {https://doi.org/10.1103/PhysRevB.43.12772}
  {\bibfield  {journal} {\bibinfo  {journal} {Phys. Rev. B}\ }\textbf {\bibinfo
  {volume} {43}},\ \bibinfo {pages} {12772} (\bibinfo {year}
  {1991})}\BibitemShut {NoStop}%
\bibitem [{\citenamefont {Goban}\ \emph {et~al.}(2014)\citenamefont {Goban},
  \citenamefont {Hung}, \citenamefont {Yu}, \citenamefont {Hood}, \citenamefont
  {Muniz}, \citenamefont {Lee}, \citenamefont {Martin}, \citenamefont
  {McClung}, \citenamefont {Choi}, \citenamefont {Chang}, \citenamefont
  {Painter},\ and\ \citenamefont {Kimble}}]{Goban2014}%
  \BibitemOpen
  \bibfield  {author} {\bibinfo {author} {\bibfnamefont {A.}~\bibnamefont
  {Goban}}, \bibinfo {author} {\bibfnamefont {C.-L.}\ \bibnamefont {Hung}},
  \bibinfo {author} {\bibfnamefont {S.-P.}\ \bibnamefont {Yu}}, \bibinfo
  {author} {\bibfnamefont {J.}~\bibnamefont {Hood}}, \bibinfo {author}
  {\bibfnamefont {J.}~\bibnamefont {Muniz}}, \bibinfo {author} {\bibfnamefont
  {J.}~\bibnamefont {Lee}}, \bibinfo {author} {\bibfnamefont {M.}~\bibnamefont
  {Martin}}, \bibinfo {author} {\bibfnamefont {A.}~\bibnamefont {McClung}},
  \bibinfo {author} {\bibfnamefont {K.}~\bibnamefont {Choi}}, \bibinfo {author}
  {\bibfnamefont {D.}~\bibnamefont {Chang}}, \bibinfo {author} {\bibfnamefont
  {O.}~\bibnamefont {Painter}},\ and\ \bibinfo {author} {\bibfnamefont
  {H.}~\bibnamefont {Kimble}},\ }\bibfield  {title} {\bibinfo {title}
  {Atom–light interactions in photonic crystals},\ }\href
  {https://doi.org/10.1038/ncomms4808} {\bibfield  {journal} {\bibinfo
  {journal} {Nat. Commun.}\ }\textbf {\bibinfo {volume} {5}},\ \bibinfo {pages}
  {3808} (\bibinfo {year} {2014})}\BibitemShut {NoStop}%
\bibitem [{\citenamefont {Lambropoulos}\ \emph {et~al.}(2000)\citenamefont
  {Lambropoulos}, \citenamefont {Nikolopoulos}, \citenamefont {Nielsen},\ and\
  \citenamefont {Bay}}]{Lambropoulos_2000}%
  \BibitemOpen
  \bibfield  {author} {\bibinfo {author} {\bibfnamefont {P.}~\bibnamefont
  {Lambropoulos}}, \bibinfo {author} {\bibfnamefont {G.~M.}\ \bibnamefont
  {Nikolopoulos}}, \bibinfo {author} {\bibfnamefont {T.~R.}\ \bibnamefont
  {Nielsen}},\ and\ \bibinfo {author} {\bibfnamefont {S.}~\bibnamefont {Bay}},\
  }\bibfield  {title} {\bibinfo {title} {Fundamental quantum optics in
  structured reservoirs},\ }\href {https://doi.org/10.1088/0034-4885/63/4/201}
  {\bibfield  {journal} {\bibinfo  {journal} {Rep. Prog. Phys.}\ }\textbf
  {\bibinfo {volume} {63}},\ \bibinfo {pages} {455} (\bibinfo {year}
  {2000})}\BibitemShut {NoStop}%
\bibitem [{\citenamefont {Gonz\'alez-Tudela}\ and\ \citenamefont
  {Porras}(2013)}]{PhysRevLett.110.080502}%
  \BibitemOpen
  \bibfield  {author} {\bibinfo {author} {\bibfnamefont {A.}~\bibnamefont
  {Gonz\'alez-Tudela}}\ and\ \bibinfo {author} {\bibfnamefont {D.}~\bibnamefont
  {Porras}},\ }\bibfield  {title} {\bibinfo {title} {Mesoscopic entanglement
  induced by spontaneous emission in solid-state quantum optics},\ }\href
  {https://doi.org/10.1103/PhysRevLett.110.080502} {\bibfield  {journal}
  {\bibinfo  {journal} {Phys. Rev. Lett.}\ }\textbf {\bibinfo {volume} {110}},\
  \bibinfo {pages} {080502} (\bibinfo {year} {2013})}\BibitemShut {NoStop}%
\bibitem [{\citenamefont {Rabl}\ \emph {et~al.}(2010)\citenamefont {Rabl},
  \citenamefont {Kolkowitz}, \citenamefont {Koppens}, \citenamefont {Harris},
  \citenamefont {Zoller},\ and\ \citenamefont {Lukin}}]{Rabl2010}%
  \BibitemOpen
  \bibfield  {author} {\bibinfo {author} {\bibfnamefont {P.}~\bibnamefont
  {Rabl}}, \bibinfo {author} {\bibfnamefont {S.~J.}\ \bibnamefont {Kolkowitz}},
  \bibinfo {author} {\bibfnamefont {F.~H.~L.}\ \bibnamefont {Koppens}},
  \bibinfo {author} {\bibfnamefont {J.~G.~E.}\ \bibnamefont {Harris}}, \bibinfo
  {author} {\bibfnamefont {P.}~\bibnamefont {Zoller}},\ and\ \bibinfo {author}
  {\bibfnamefont {M.~D.}\ \bibnamefont {Lukin}},\ }\bibfield  {title} {\bibinfo
  {title} {A quantum spin transducer based on nanoelectromechanical resonator
  arrays},\ }\href {https://doi.org/10.1038/nphys1679} {\bibfield  {journal}
  {\bibinfo  {journal} {Nat. Phys.}\ }\textbf {\bibinfo {volume} {6}},\
  \bibinfo {pages} {602} (\bibinfo {year} {2010})}\BibitemShut {NoStop}%
\end{thebibliography}

\end{document}